\documentclass[aps,prl,amsmath,amsfonts,amssymb,notitlepage,twocolumn,superscriptaddress,footinbib,floatfix]{revtex4-2}
\usepackage{CJK}
\usepackage{bm}
\usepackage{soul}
\usepackage[caption=false,subrefformat=parens,labelformat=parens]{subfig}
\usepackage{color} 
\usepackage[table,dvipsnames]{xcolor}
\usepackage[colorlinks,linktocpage,bookmarks=false]{hyperref}
\usepackage{enumitem}
\usepackage{ulem}
\usepackage{xspace}

\newcommand\myshade{85}
\definecolor{myrulecolor}{RGB}{150,20,0}
\colorlet{mylinkcolor}{violet}
\colorlet{mycitecolor}{YellowOrange}
\colorlet{myurlcolor}{Aquamarine}
\hypersetup{
	linkcolor = myurlcolor!\myshade!black,
	citecolor = mycitecolor!\myshade!black,
	urlcolor = myrulecolor!\myshade!black,
}  

\usepackage{graphicx}
\graphicspath{{FIG/}}
\usepackage{multirow}
\usepackage{braket} 
\usepackage{booktabs} 

\AtBeginDocument{
	\heavyrulewidth=.08em
	\lightrulewidth=.05em
	\cmidrulewidth=.03em
	\belowrulesep=.65ex
	\belowbottomsep=0pt
	\aboverulesep=.4ex
	\abovetopsep=0pt
	\cmidrulesep=\doublerulesep
	\cmidrulekern=.5em
	\defaultaddspace=.5em
}

\newcommand{\beq}{\begin{equation}}
	\newcommand{\eeq}{\end{equation}}
\newcommand{\bea}{\begin{eqnarray}}
	\newcommand{\eea}{\end{eqnarray}}

\renewcommand\[{\begin{equation}}
	\renewcommand\]{\end{equation}}

\newcommand{\padic}{$p$-adic}

\newcommand*\Laplace{\mathop{}\!\mathbin\bigtriangleup}

\makeindex

\begin{document} 
	\begin{CJK*}{UTF8}{gbsn} 
		\title{$p$-adic Holography from the  Hyperbolic Fracton Model}
		\author{Han Yan (闫寒)}
        \affiliation{Department of Physics and Astronomy, Rice University, Houston, TX 77005, USA}
		\affiliation{Smalley-Curl Institute, Rice University, Houston, TX 77005, USA}
		\author{Christian B. Jepsen}
		\affiliation{Simons Center for Geometry and Physics, SUNY, Stony Brook, NY 11794, USA}
		\author{Yaron Oz}  
		\affiliation{Raymond and Beverly Sackler School of Physics and Astronomy, Tel-Aviv University, Tel-Aviv 69978, Israel}
\begin{abstract}
We reveal a low-temperature duality between the hyperbolic lattice model featuring fractons and infinite decoupled copies of Zabrodin's $p$-adic model of AdS/CFT. The core of the duality is the subsystem symmetries of the hyperbolic fracton model, which always act on both the boundary and the bulk. These subsystem symmetries are associated with fractal trees embedded in the hyperbolic lattice, which have the same geometry as Zabrodin's model. The fracton model, rewritten as electrostatics theory on these trees, matches the equation of motion of Zabrodin's model. The duality extends from the action to lattice defects as $p$-adic black holes.
\end{abstract}
\maketitle
\end{CJK*}
	
	\noindent\textbf{\textit{Introduction.  --  }}  
	The contemporary landscape of theoretical physics is marked by an exhilarating fusion of quantum many-body systems, quantum gravity, and quantum information,   
	guided by the holographic principle and the anti-de Sitter/conformal field theory (AdS/CFT) correspondence \cite{Maldacena1999,Witten1998}. This 
 duality not only provides a profound insight into quantum gravity but also offers a robust tool for addressing condensed matter problems \cite{Zaanen2015_ADSCMTbook,Hartnoll:2018xxg}.
	Additionally, the development of numerous tensor-network holographic toy models reveals the quantum-informational correcting feature of holographic entanglement \cite{Swingle2012,Pastawski2015,Yang2016}.
	
	Fracton states of matter \cite{ChamonPhysRevLett.94.040402,YoshidaPhysRevB.88.125122,BRAVYI2011839,Haah2011,Vijay2015,Pretko2017a,Nandkishoreannurev,pretko2020fractonReview,gromov2022fractonReview}, closely tied with Lifshitz gravity \cite{Xu2010PhysRevD,Pretko2017PhysRevD}, have recently been incorporated into the AdS/CFT landscape.  
	Prior investigations \cite{Yan2019PhysRevBfracton1,Yan2019PhysRevBfracton2,Yan2020PhysRevB} have indeed revealed that various holographic information properties of fracton models in hyperbolic space are consistent, akin to holographic states built via tensor networks \cite{Pastawski2015,Yang2016,Jahneaaw0092}. 
	However, despite these advancements, essential questions regarding the effective AdS bulk theory and corresponding boundary CFT for these states remain a mystery.

	In this work, we help answer these questions by establishing a connection between the \padic~Zabrodin model \cite{zabrodin1989non}, which is arguably the simplest toy model of AdS/CFT, and   hyperbolic lattice models featuring fractons. 
	Notably, the lattice model's subsystem symmetry is described by fractal trees identical to the $(p+1)$-fractal tree in the Zabrodin model, forming the basis for  the connection. 
	The connection between the two models can be established from the action and information properties to \padic~black holes (BHs) as lattice defects.
	This discovery builds a valuable framework for understanding physics on both sides of the correspondence.
	Furthermore, it enriches the AdS/condensed matter theory (CMT) program \cite{Zaanen2015_ADSCMTbook,Hartnoll:2018xxg,breuckmann2020critical} by presenting an example of condensed matter theory within the bulk, rather than conventionally on the boundary. \\

\begin{figure}[ht!]
		\centering    
		\subfloat[\label{fig_E_subsymetry}]{\includegraphics[width=0.5\columnwidth]{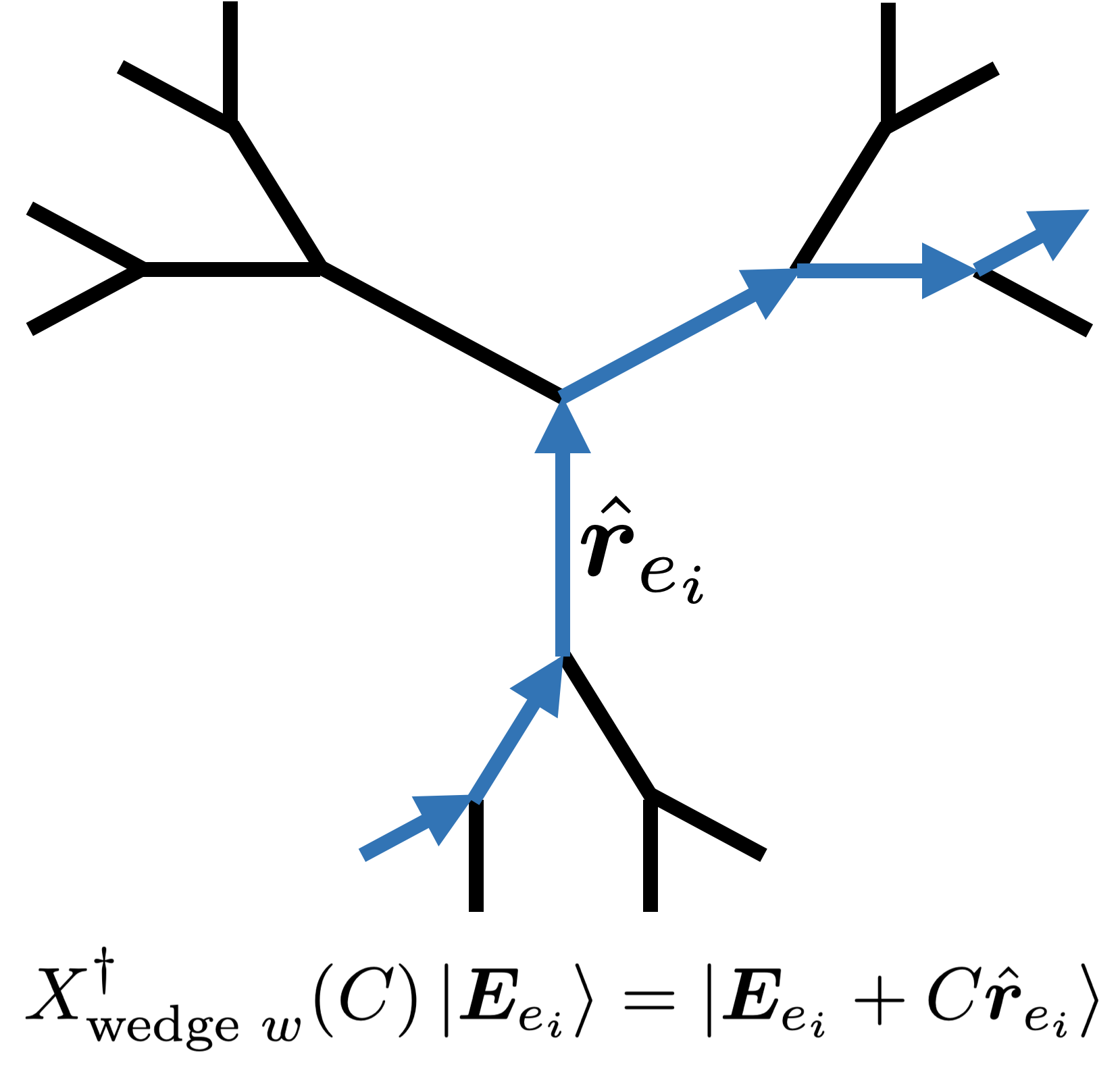}}
		\subfloat[\label{fig_rindler_1panel}]{\includegraphics[width=0.47\columnwidth]{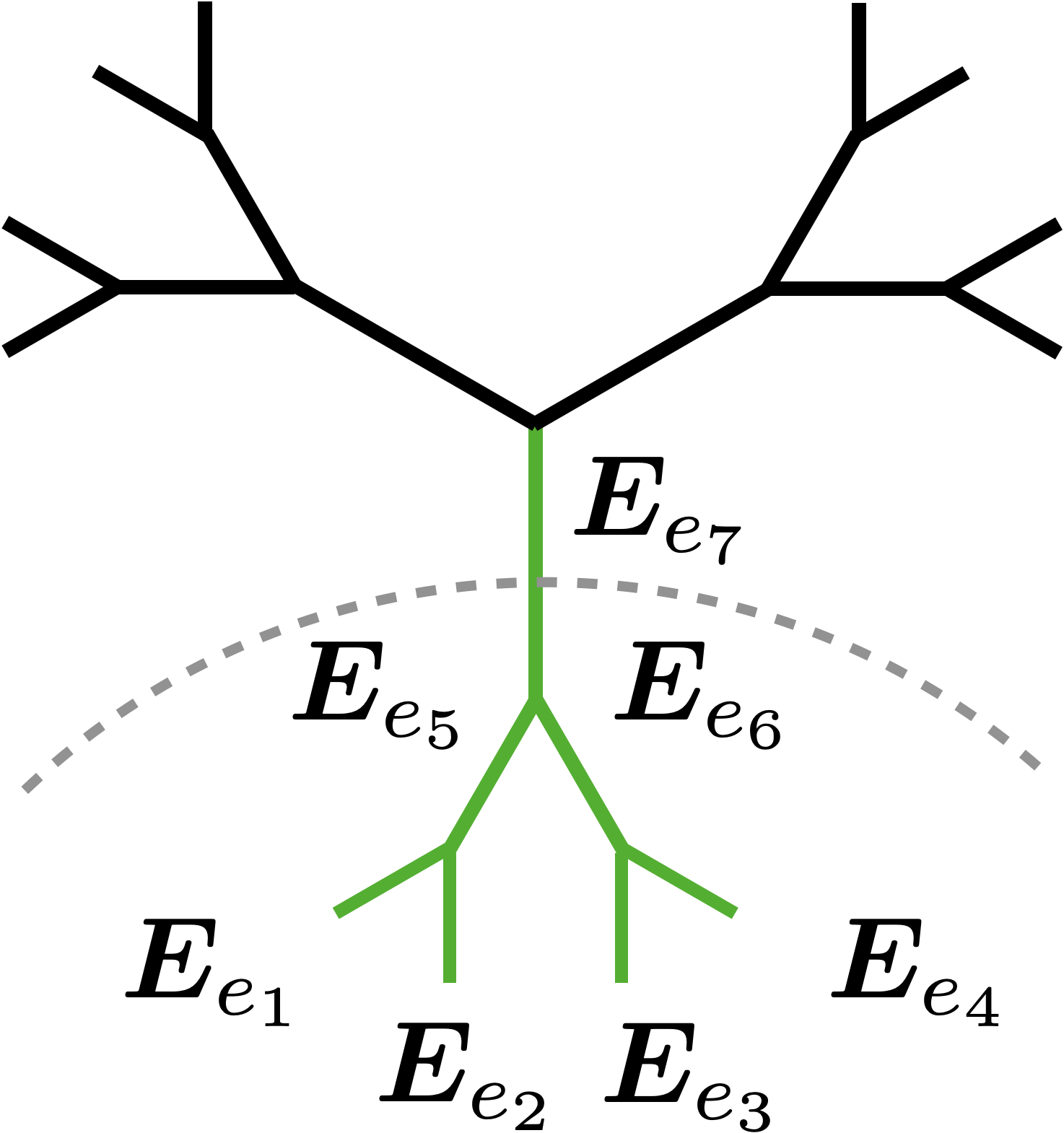}}	 
		\caption{\label{fig_E_subsymetry_big} 
			(a) Subsystem symmetry in the dual model as injecting a flux of electric field.	(b)
			AdS-Rindler reconstruction: knowing boundary electric fields $E_1$ to $E_4$ allows reconstruction up to $E_7$ (minimal covering surface), but not beyond.}   
	\end{figure}

	\noindent\textbf{\textit{Zabrodin's model.  --  }} Zabrodin's \padic~ model \cite{zabrodin1989non}, a simple AdS/CFT toy model, uses Gaussian fields on the vertices of an infinite $(p+1)$-regular tree $\mathcal{T}_p$ to discretize the AdS bulk. This tree is also known as the Bruhat-Tits tree or $(p+1)$-degree fractal tree, as shown in Fig.~\ref{fig_E_subsymetry} for a $3$-regular tree.  
	Its action is given by
	\begin{align}
		\label{eq:Sbulk}
		S_{\text{bulk}}
		=\frac{\gamma}{2}\sum_{v\in \mathcal{T}_p}\sum_{v'\sim v}
		(\phi_v-\phi_{v'})^2\,,
	\end{align}
	where $v'$ denotes neighbor vertices of $v$. The model's equation of motion is given in terms of a graph Laplacian:
	\[\label{EQN_Zabrodin_EOM}
	\Laplace \phi_{v} \equiv (p+1) \phi_{v } - \sum_{v' \sim v}\phi_{v'} = 0 .
	\] 
	Focusing on prime $p$, Zabrodin established a duality between this discrete theory and a continuous boundary theory,
	\begin{align}
		\label{eq:Sboundary}
		S_{\text{boundary}}
		=\frac{\gamma}{4}\frac{p(p-1)}{p+1}\int_{\mathbb{Q}_p}dx\,dy\,\frac{(\phi(x)-\phi(y))^2}{|x-y|_p^2}\,,
	\end{align}
	where $\mathbb{Q}_p$ is the field of $p$-adic numbers and $|\cdot |_p$ is the $p$-adic norm \cite{koblitz,gouvea}. In the limit, where source terms in \eqref{eq:Sbulk} approach the tree boundary, the partition functions of \eqref{eq:Sbulk} and \eqref{eq:Sboundary} become identical. 
	The theory \eqref{eq:Sboundary} had previously been introduced as an effective theory for the endpoints of $p$-adic strings, from which the $p$-adic   Veneziano amplitude \cite{freund1987non} can be derived  \cite{spokoiny1988quantum,zhang1988lagrangian}. 
	With his model \eqref{eq:Sboundary}, Zabrodin had discovered the worldsheet theory of $p$-adic strings. 
	With the advent of $p$-adic AdS/CFT  \cite{gubser2017p,heydeman2016tensor}, Zabrodin's duality was reinterpreted as the first example of $p$-adic holography.
	From this new perspective, the tree $\mathcal{T}_p$ is instead viewed as  the $p$-adic version of AdS bulk, and the $p$-adic numbers is  the boundary space in which $p$-adic CFTs live. \\
	
\begin{figure}[ht!]
		\centering
		\includegraphics[width=0.5\columnwidth]{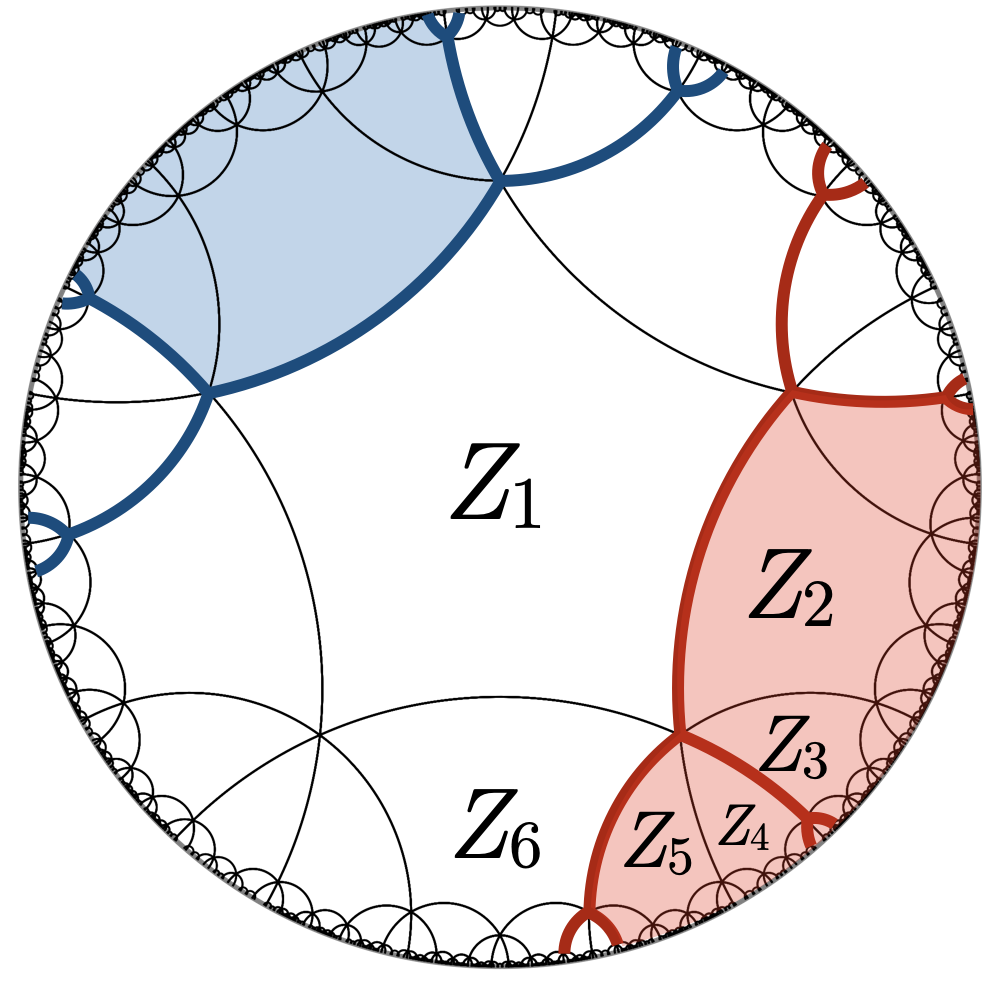}
		\caption{\label{Fig_hyperbolic} 
			$(5,6)$ tessellation of the hyperbolic plane on  the Poincar\'e disk:
			all polygons on the disk are identical, but look smaller when drawn farther from the center. 
			The thick blue and red lines are fractal trees, and the shaded regions are associated with different subsystem symmetries.
		} 
	\end{figure}

	\noindent\textbf{\textit{Hyperbolic fracton model with Ising perturbations. -- }}
	In this work, we discuss the Hyperbolic Fracton Model (HFM)   with ferromagnetic Ising interactions (HFM+Ising). These models are placed on hyperbolic lattices, tessellations of a 2D plane with constant negative curvature by polygons. The lattices are characterized by Schl\"{a}fli symbols, pairs of integers $(m,n)$, which represent tessellations of $m$-sided regular polygons with $n$ polygons meeting at each vertex. This configuration necessitates the condition $m^{-1}+{n}^{-1}<1/2$, permitting infinitely many different hyperbolic tessellations. An Example with Schl\"{a}fli symbol $(5,6)$ are shown in Fig.~\ref{Fig_hyperbolic}. Our study is primarily concerned with tessellations having an even $n$, and uses the $(5,6)$ lattice as the concrete example. 
	
	In the HFM, the scalar degrees of freedom $Z_{p_i}$ are located at the centers of  polygons labeled by $p_i$'s.
	The total Hamiltonian is given by 
	\[
	\mathcal{H}_\text{HFM-Ising} = \mathcal{H}_\text{HFM} + \mathcal{H}_\text{Ising} .
	\] 
	It includes the HFM part $\mathcal{H}_\text{HFM}$ and the Ising part $\mathcal{H}_\text{Ising}$.
	The Ising part is the familiar Ising model with ferromagnetic nearest-neighbor interactions, 
	\[ \label{EQN_Ising_Ham}
	\mathcal{H}_\text{Ising} = \alpha \sum_{\langle i, j \rangle} (Z_{p_i} - Z_{p_j})^2.
	\] 
	We now elaborate on the HFM  component, $\mathcal{H}_\text{HFM}$. The HFM extends the model initially presented in Refs.~\cite{Yan2019PhysRevBfracton1,Yan2019PhysRevBfracton2,Yan2020PhysRevB}, with excitations analogous to fractons in gapped fracton orders. Its Hamiltonian is:
	\begin{equation}
		\label{EQN_HFM_general}
		\mathcal{H}_\text{HFM} = U \sum_{v \in \text{vertices}}\left(Z_{p_1} - Z_{p_2}+  \dots -  Z_{p_n}\right)^2,
	\end{equation}
	where $p_1,\dots,p_n$ indicate the $n$ sites encircling vertex $v$ in a clockwise sequence (refer to Fig.~\ref{Fig_hyperbolic} for an example). 
	
	Alternatively, considering $Z_{p_i}$ as $\mathbb{Z}_N$ numbers, one can replace $(Z_{p_1} - Z_{p_2}+ \dots - Z_{p_n})^2$ with $(\frac{N}{\pi})^2\sin^2\left[\frac{\pi}{N}(Z_{p_1} - Z_{p_2}+ \dots - Z_{p_n})\right]$. In the case where $Z_{p_i}$ adopts $\mathbb{Z}_2$ values, the model has been examined for the $(5,4)$ tessellation in Refs.~\cite{Yan2019PhysRevBfracton1,Yan2019PhysRevBfracton2}, representing a simplified extreme of the general HFM.
	
	A fracton is an excitation   of the vertex term. Note that by changing the value of a single $Z_{p_i}$, we create an $m$-multipole of fracton instead of dipole, which leads to the immobility of a single fracton in the system.
	This study concentrates on the low-energy sector in the limit  $U\gg \alpha$  and $U\gg T$ (the sector devoid of fracton excitations). We illustrate that the HFM+Ising model is congruent to an infinite set of \padic~Zabrodin models, further exploring its holographic information properties.\\
	
	\noindent\textbf{\textit{Emergent Zabrodin model from HFM. -- }} 
	We first examine $\mathcal{H}_\text{HFM}$ without Ising interactions, which   will be shown to match the tensionless-string limit of the Zabrodin's model. The primary result is the model's duality to multiple copies of the electrostatics theory on distinct $n/2-$degree fractal trees ($n/2$-regular trees). In the dual model, ground states on each tree  satisfy the EOM of the Zabrodin model (Eq.~\eqref{EQN_Zabrodin_EOM}). Additionally, we can verify the holographic information characteristics of the model within this limit.
	
	The HFM, described by Eq.~\eqref{EQN_HFM_general}, exhibits a large ground state degeneracy, a feature dictated by its subsystem symmetries that commute with the Hamiltonian. The key to finding them is to notice that changing an even number of neighboring variables $Z_{p_i}$ at a given vertex by the same constant $C$ leaves the vertex term invariant. 
	This results in subsystem symmetries defined by the boundaries of $n/2$-degree fractal trees, pivotal to the underlying physics. Two examples, represented by blue and red trees in thick lines, are depicted in Fig.~\ref{Fig_hyperbolic}.

	To construct these trees more formally, begin at an arbitrarily chosen vertex and select $n/2$ nonadjacent edges from the total of $n$ edges, keeping in mind that we're focusing on even-$n$ tessellations. Extend the chosen edges to the neighboring vertices and repeat the selection process, making sure to include the previously chosen edges, thus making the choice unique. Continuing this process, we construct a degree-$n/2$ fractal tree embedded within the hyperbolic lattice, extending the selected edges indefinitely. This method yields an infinitely many  trees on the lattice in the thermodynamic limit.
	
	We then define the \textit{fractal-tree wedge} as the part of the lattice bounded by the edges of a  tree, a pattern also seen in hyperbolic fracton orders \cite{yan2022ycube}.
	In Fig.~\ref{Fig_hyperbolic}, the blue and red shaded regions serve as examples.
	Given a classical state, denoted as $\prod_{p_i}\ket{Z_{p_i}}$, the associated subsystem symmetry for a particular wedge $w$ can be defined via the following operation:
	\[
	X^\dagger_{\text{wedge } w} (C) \ket{Z_{p_i}}  =   \ket{Z_{p_i} + C} , \quad \forall {p_i} \in w,
	\] 
	while $X^\dagger_{\text{wedge } w} (C)$ acts as the identity operator on $\ket{Z_{p_j}}$ for all plaquettes $p_j$ outside $w$.
	
	\begin{figure}[ht!]
		\centering    
		\subfloat[\label{Fig_E_definition}]{\includegraphics[width=0.5\columnwidth]{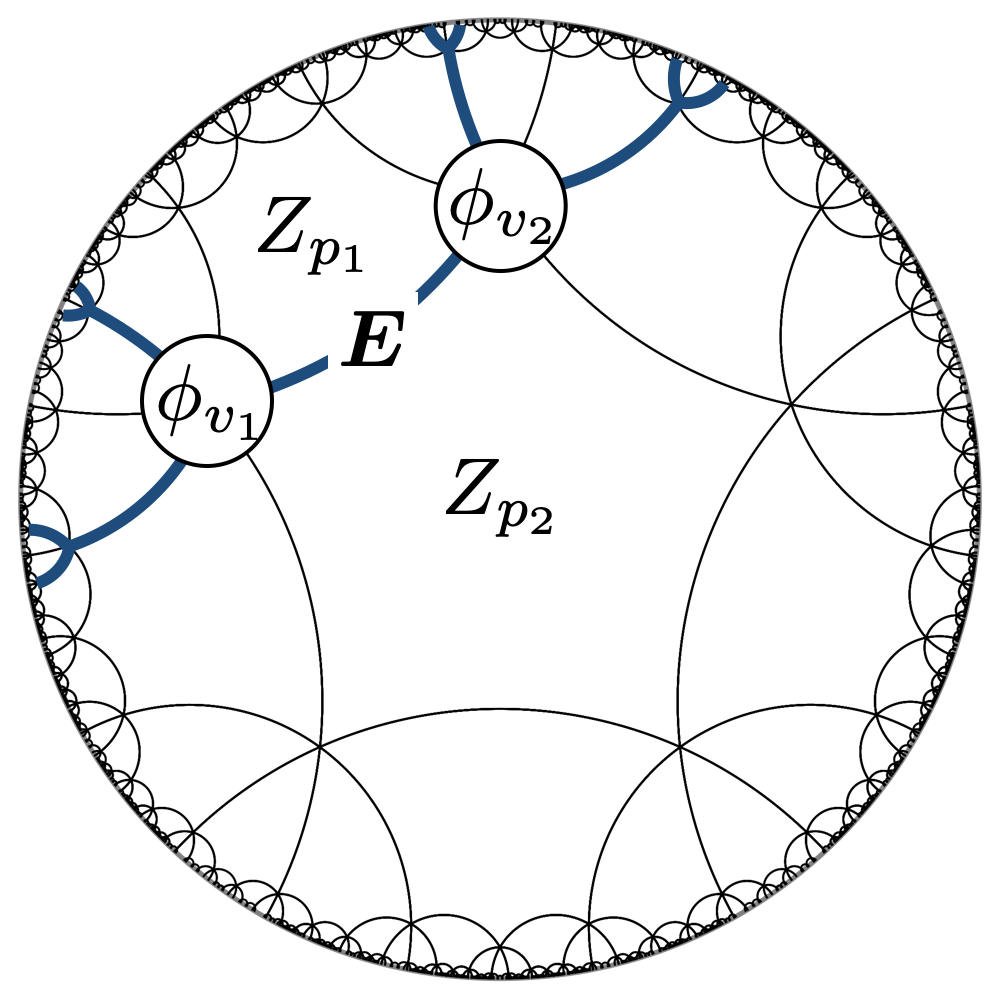}}
		\subfloat[\label{Fig_vertex_ham}]{\includegraphics[width=0.5\columnwidth]{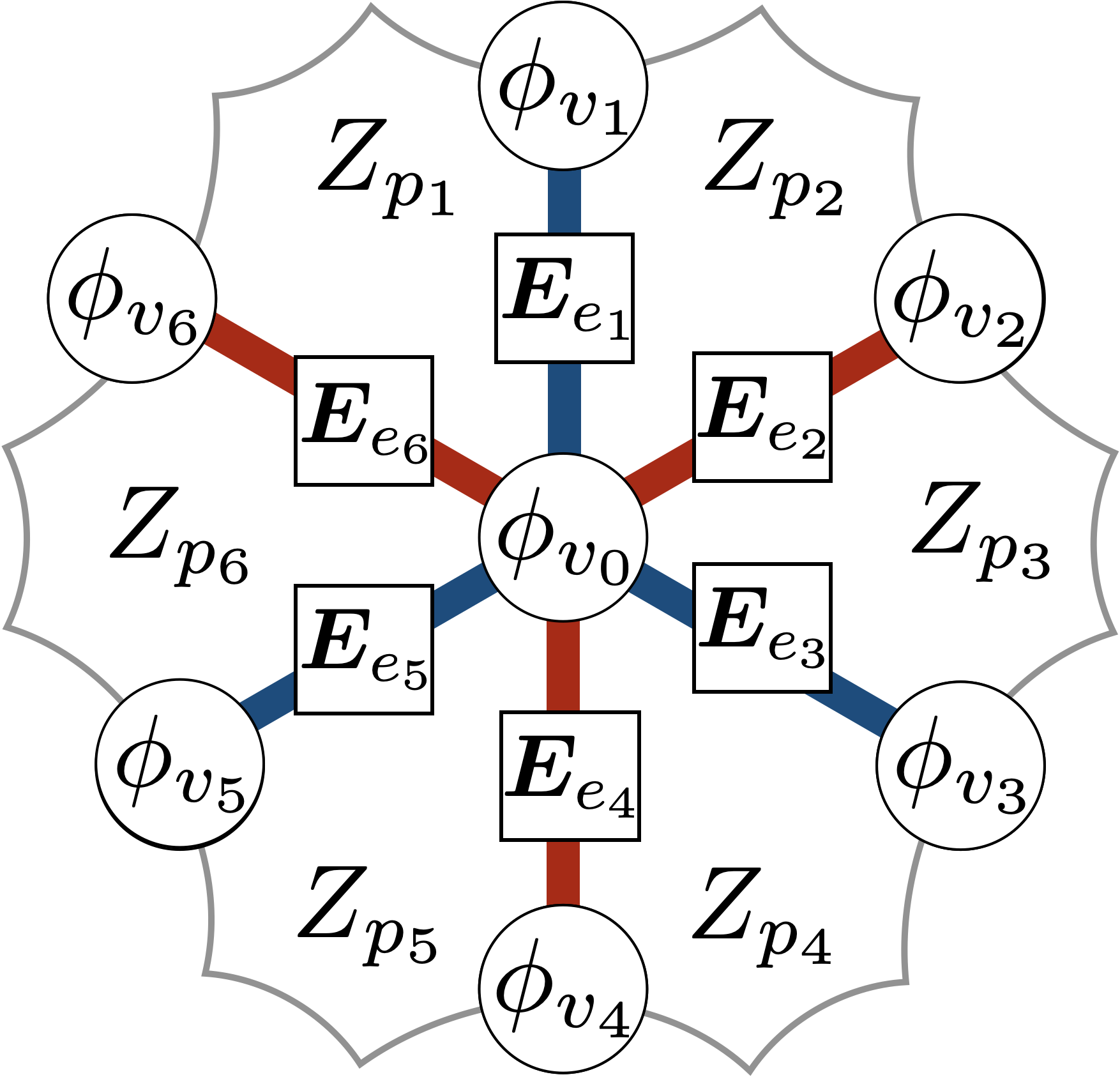}}
		\caption{
			Degree of freedom (DOF) $Z$, $\bm{E}$, and $\phi$ defined on the plaquette centers, edges, 	and vertices.
		}   
	\end{figure}

	We'll now formulate the dual model to the HFM, aiding our understanding of subsystem symmetries and illuminating its connection to the Zabrodin \padic\ model.

	We start by defining 1$d$ vector DOFs, denoted as $\bm{E}_{e_i}$, which reside on the lattice's edges, $e_i$, and point along them. Each edge, $e$, can be described in two ways: it's sandwiched by the plaquettes $p_i,\ p_j$, denoted as $e = p_{i,j}$, and it has vertices $v_k,\ v_l$ at its endpoints, denoted as $e = v_{k,l}$.
	For an edge as depicted in Fig.~\ref{Fig_E_definition}, we define:
	\[
	\bm{E}_{v_{1,2}} \equiv (Z_{p_1} - Z_{p_2})  \hat{\bm{r}}_{v_{1,2}}  = (-Z_{p_1} + Z_{p_2})   \hat{\bm{r}}_{v_{2,1}}  =   \bm{E}_{v_{2,1}} .
	\]
	Note that the order $v_1\rightarrow p_1 \rightarrow v_2 \rightarrow p_2$ around the edge is in a clockwise direction (see Appendix for the counting of DOFs).

	At a  vertex, the term $ (Z_{p_1} - Z_{p_2}+  \dots -  Z_{p_q})^2$ can be recast in terms of $\bm{E}$ as illustrated in Fig.~\ref{Fig_vertex_ham} for the special case $q=6$,
	\[
	\label{EQN_Z_E_div_relation}
	\begin{split}
		&  (Z_{p_1} - Z_{p_2}+  \dots -  Z_{p_6})^2 \\
		= &(\bm{E}_{v_{0,1}} \cdot \hat{\bm{r}}_{v_{0,1}} +\bm{E}_{v_{0,3}} \cdot \hat{\bm{r}}_{v_{0,3}}+\bm{E}_{v_{0,5}} \cdot \hat{\bm{r}}_{v_{0,5}})^2  \\
		=   & (\nabla \cdot \bm{E})^2 \text{ on the three blue edges}
	\end{split}
	\] 
	Notably, the $\bm{E}$ fields associated with an $n/2$-degree fractal tree correlate, while remaining uncoupled from  DOFs on other tree, except for an exact inter-tree constraint at each vertex,
	\[
	\label{EQN_two_tree_one_vertex}
	(\nabla \cdot \bm{E})^2 \text{ on   blue edges} = (\nabla \cdot \bm{E})^2 \text{ on red   edges}.
	\]
But in the low-energy sector where fracton excitations are absent, these constraints no longer couple the fractal trees to each other.  
	
	Let's delve into the physics on a single fractal tree.
	The Hamiltonian on the tree  of interest is  
	\[
	\label{EQN_fractal_tree_electro}
	\mathcal{H}_\text{tree}  =  U\sum_{v_i  \text{  on } \mathcal{T}} (\nabla \cdot \bm{E})^2 . 
	\]
	This Hamiltonian unveils that the model equates to an electrostatics theory on the tree, involving solely the electric field sector of electrodynamics. It imposes an energy cost on charge excitations, thus the low-energy sector corresponds to charge-free electric field configurations.
	The subsystem symmetry in the dual model is expressed as: 
	\[
	\begin{split}
		&X^\dagger_{\text{wedge } w} (C)  \ket{\bm{E}_{e_i}}  = \ket{\bm{E}_{e_i} + C\hat{\bm{r}}_{e_i}}  ,\\
		&\quad \text{ for edges $e_i$ on the boundary of  wedge $w$.}
	\end{split}
	\]
	In the context of electrostatics, this has a clear interpretation: the injection of an electric field string from one end on the boundary  to the other. Given the absence of loops on a fractal tree, this is the sole method of flux injection without charge creation, as depicted in Fig.~\ref{fig_E_subsymetry}.
	Thus we identify a set of multiple copies of electrostatics on $n/2$-degree fractal trees as a dual model to the HFM. 
	
	This identification   aids in elucidating certain holographic properties such as the AdS-Rindler reconstruction.
	Using Fig.~\ref{fig_rindler_1panel} as an example, an observer measuring  boundary electric fields $E_1$ to $E_4$ can  reconstruct bulk DOFs up to $E_7$ (minimal covering surface) using the charge-free condition at every vertex, but not beyond (see Appendix for details).
	
	Furthermore, we can recast the fractal-tree electrostatics using the electric field potential,
	whose low energy sector explicitly obeys the EOM of  the Zabrodin \padic\ model.
	We assign electric potential $\phi_{v_i}$'s to the  vertices.
	The electric field between   vertices $v_1,\ v_2$ is   $\bm{E}_{v_{1,2}}  =  (\phi_{v_1} - \phi_{v_2} )\hat{\bm{r}}_{v_{1,2}}$.
	The Hamiltonian~\eqref{EQN_fractal_tree_electro}   becomes 
	\[
	\label{EQN_tree_map_to_Z_model}
	\begin{split}
		\mathcal{H}_\text{tree} & = U\sum_{v_i  \text{  on } \mathcal{T}} \left( (\frac{n}{2}  )\phi_{v_i} - \phi_{v_{i,1}} - \dots - \phi_{v_{i,n/2}} \right)^2  \\
		&\equiv   U\sum_{v_i  \text{  on } \mathcal{T}} \left( \Laplace\phi \right)^2.
	\end{split}
	\]
	Here, $v_{i,j}$ denote the $n/2$ vertices next to $v_i$, and $\Laplace$ denotes the Laplacian operator on the tree. 
	Hence, the ground states of the dual model to HFM are precisely the solutions to the EOM of the Zabrodin model (Eq.~\eqref{EQN_Zabrodin_EOM}), with the prime $p$ of Zabrodin's model being related to the second integer $n$ of the Schl\"{a}fli symbol via the relation $n=2p+2$. In the case when $(n-2)/2$ is not a prime, the bulk HFM does not   change behaviour substantially, but the dual tree model no longer admits a convenient holographic description in terms of $p$-adic numbers; what is lost is the multiplicative property of the norm: $|ab|_p=|a|_p|b|_p$.
	
	Note that the degenerate ground states of $\mathcal{H}_\text{HFM}$ contribute unequally to the action of the Zabrodin model, (Eq.~\eqref{eq:Sbulk}) 
	\[  
	S_{\text{bulk}}
	=\frac{\gamma}{2}\sum_{v\in \mathcal{T}_{n/2}}\sum_{v'\sim v}
	(\phi_v-\phi_{v'})^2 = \gamma \sum_{e \in \mathcal{T}_{n/2}} {\bm{E}_e}^2,
	\] 
	which differentiates the   charge-free electric field configurations by the familiar $\gamma {\bm{E} }^2$ term, instilling tension to the strings on the fractal tree.
	Consequently,   the pure HFM model realizes the Zabrodin model's tensionless-string limit.
	
	To account for the omitted string tension contribution, we should enable the Ising sector $\mathcal{H}_\text{Ising}$  (Eq.~\eqref{EQN_Ising_Ham}) and  match $\alpha = \gamma$, so that
	\[
	\alpha (Z_{p_i} - Z_{p_j}) ^2 =\gamma  {\bm{E} }_{e={p_{i,j}}}^2.
	\]
	Once the Ising Hamiltonian is turned on, the Hamiltonian associated to each of the trees exactly matches that of Zabrodin's model. An exact duality is not present generically, however, owing to the fact that two trees intersect at each vertex on the lattice, and the edge degrees of freedom of these trees are constrained to produce identical vertex terms. Only in the limit when $U$ is much greater than $kT=\beta^{-1}$, do the trees decouple, as the vertex terms are all required to assume their lowest possible value. This is a low-temperature limit according to the scale set by $U$, but not according to the scale set by $\alpha$: we do not assume $\alpha>> kT$.

	We see here that the large $U$ limit of the HFM+Ising in its action on a single fractal tree, as an electrostatics theory, is equivalent to the Zabrodin model. Consequently, the HFM+Ising model on the full hyperbolic lattice, as infinitely many fractal trees defining the subsystem symmetries,	is a model hosting infinitely many  holographic \padic~Zabrodin models.\\ 
	
	\noindent\textbf{\textit{Lattice defects as \padic~black holes. --}}
	Finally, we explain how lattice defects on HFM corresponds to BTZ black holes  in \padic~AdS/CFT. We set $\alpha=0$ and turn off the Ising interaction, because in this case the electrostatics picture of the BH offers a particularly simple intuitive understanding of the microstates and entropy of the BH. 
	
\begin{figure}[th]
		\includegraphics[width=0.8\columnwidth]{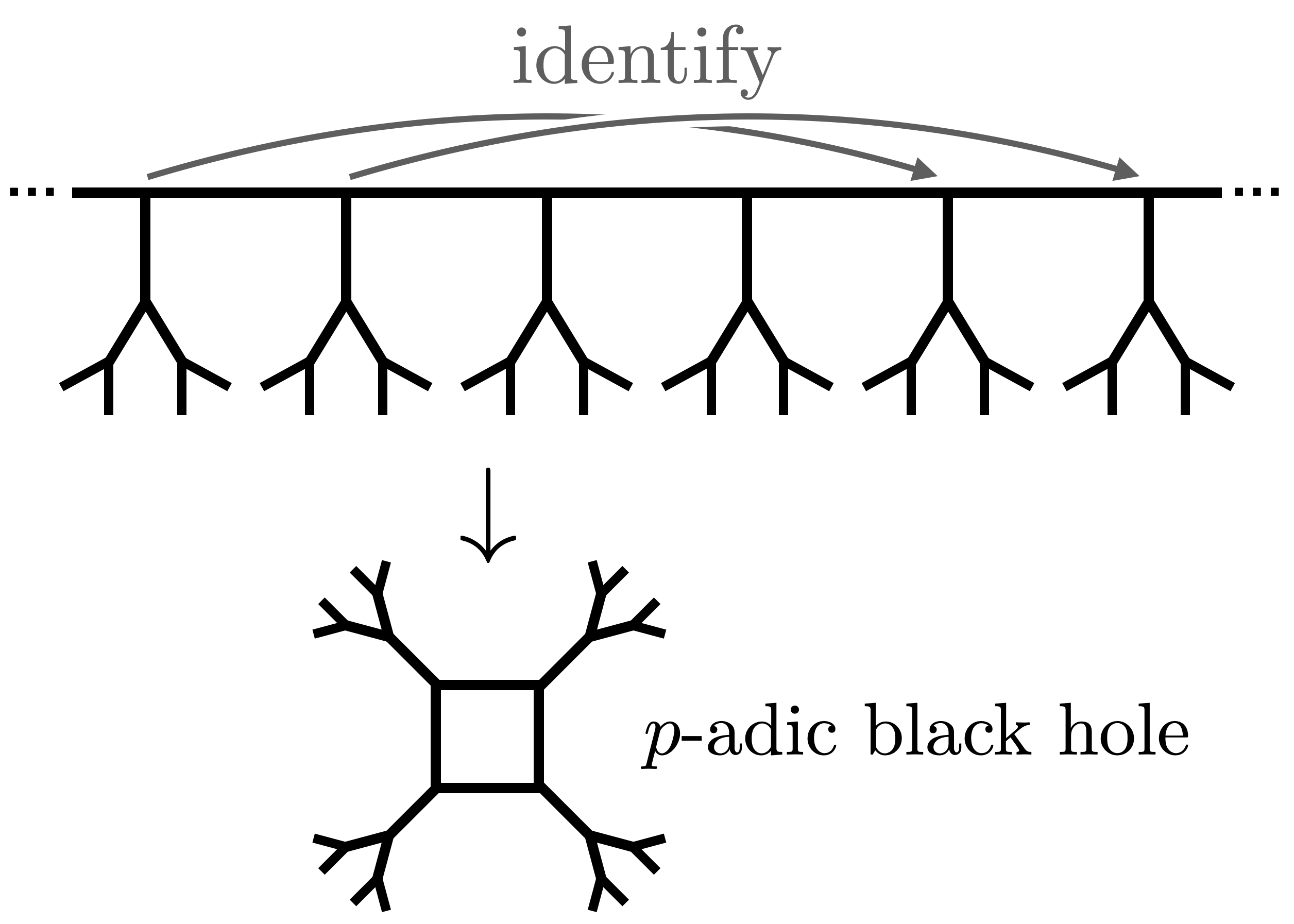}  
		\caption{Quotient construction of $p$-adic BTZ black hole. Top: Picking any doubly-infinite path in the Bruhat-Tits tree, the shifting of all vertices by $L$ edges along this path is an isometry for any $L\in \mathbb{N}$. Bottom: Identifying vertices related by such an isometry results in the $p$-adic BTZ black hole. In this case $L=4$ and $p=2$. 
			\label{padicBTZ}} 
	\end{figure}

	Defined on a rigid lattice, the  \padic~model  has no notion of dynamical gravity. 
	But Refs.~\cite{manin2002holography,heydeman2016tensor} 
	proposed the notion of $p$-adic BHs as topological objects, constructed in analogy with the arithmetic interpretation of the BTZ BH \cite{banados1992black} as a quotient of the isometry group of AdS by a discrete subgroup \cite{steif1996supergeometry}. 
	The $p$-adic BTZ BH boils down to the same action defined on  
	a  graph with a loop of $L$ edges and $L$ vertices,  attached to $L$ fractal trees of coordination number $(p+1)$, see Fig.~\ref{padicBTZ}. 
	Subsequent papers \cite{heydeman2018nonarchimedean,hung2019p,ebert2019probing,chen2021bending1,chen2021bending2} have corroborated the identification of this graph with a type of BTZ BH, with $L$ as the size of the BH perimeter. 
	
\begin{figure}[th]
		\centering    \includegraphics[width=\columnwidth]{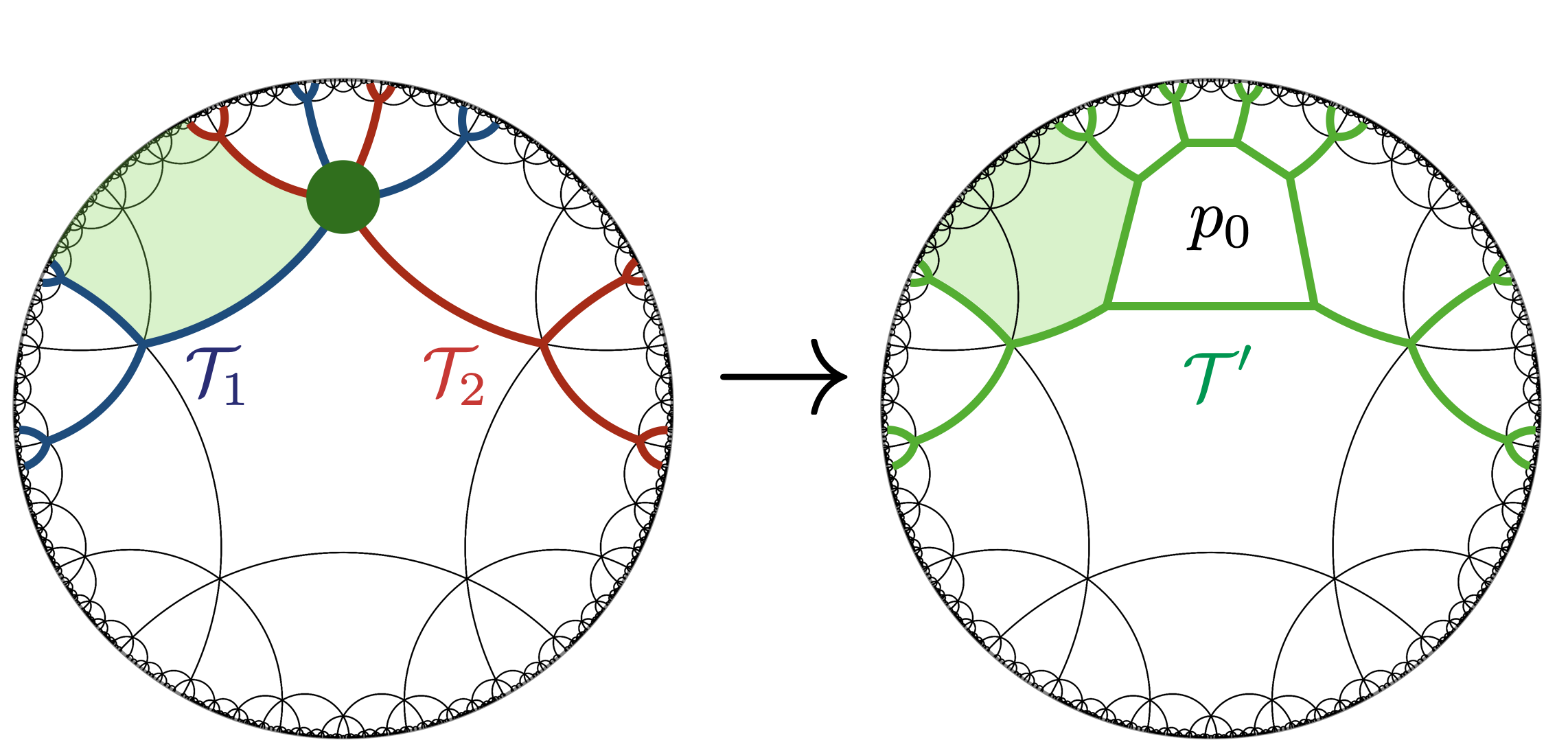}
		\caption{\label{Fig_fracton_bh}  
			A defect on the hyperbolic lattice and the merger of two fractal trees that it engenders.
		} 
	\end{figure}

	We  now explain how a defect in the HFM can lead to the emergence of a \padic\ BH. The defect sits at a vertex, as shown by the green disk in Fig.~\ref{Fig_fracton_bh}, and takes the form of a new plaquette with DOF $Z_{p_0}$ there.  
	
	Initially, the vertex has two trees, $\mathcal{T}_1$ and $\mathcal{T}_2$, whose respective wedges determine the subsystem symmetries of the HFM model. However, the inclusion of the defect merges $\mathcal{T}_1$ and $\mathcal{T}_2$ into a single tree $\mathcal{T}'$, with a loop at its center, as shown in Fig.~\ref{Fig_fracton_bh}. 
	Every wedge bounded by the new tree $\mathcal{T}'$    defines  a subsystem symmetry. The low-energy domain of the electrostatics theory living on $\mathcal{T}'$ abides by the EOM for the \padic~theory  with the BTZ BH. With the Ising term included, it also matches the action.
	
	The introduction of a defect and  suppression of the vertex term leads to an increase in the ground state degeneracy, corresponding to the microstates of the $p$-adic BH and its resultant entropy. Consider the simplest BH from a single defect, depicted in Fig.~\ref{Fig_fracton_bh}. Initially, the electric fields on trees $\mathcal{T}_1$ and $\mathcal{T}_2$ comply with individual charge-free conditions, $\rho_1 = 0$ and $\rho_2 = 0$ if one measured the electric fields on the boundary of a region containing the BH. After merging these into a single tree $\mathcal{T}'$ (or equivalently  $\mathcal{T}_1 \cup \mathcal{T}_2$ for observers outside), only a single charge-free condition applies to the whole tree: $\rho' = \rho_1 + \rho_2 = 0$. Consequently, $\rho_{1 }(= - \rho_2)$ can take non-zero values. These are the new states affiliated with the BH. Meanwhile,  boundary observers measuring $\bm{E}$ there will struggle to extract any BH information ($\rho_{1}$) unless they simultaneously measure a large boundary region covering all ends on either original tree, $\mathcal{T}_1$ or $\mathcal{T}_2$. This concept aligns with physics discussed in simpler cases where trees represent geodesics, as mentioned in Refs.~\cite{Yan2019PhysRevBfracton1,Yan2019PhysRevBfracton2}. 
	
	A defect encompassing a larger region merges $N$ trees, with a longer loop connecting all branches outside the loop. Here, the original $N$ conditions $\rho_i = 0 \ \forall\ i=1,\dots ,N$ on the ground state are reduced into a single one, $\sum_i \rho_i = 0$. 
	So the  BH  DOF is then  labeled by  a series of numbers
	$\rho_1,\ \rho_2,\ \dots,\ \rho_{N-1}$.
	The count of $N$ corresponds to the count of trees partially overlapping with the defect region.
	In the Appendix, we show that for a region of large radius (number of layers) $R$, the number of BH DOF  is
	\[ \label{EQN_BH_DOF}
	N_\text{BH}   \sim e^{ R \log \lambda_+} \sim \sinh{(R \log \lambda_+)},
	\]
	where $\lambda_+$ is a constant depending on the lattice tessellation $(m,n)$.
	The 
 number of DOF $N_\text{BH}$, which measures the entropy of the BH,  is proportional to its   horizon size.\\ 
	
	\noindent\textbf{\textit{Discussion. --  }}
	We've demonstrated that, at low temperatures, the simplest HFM coupled with weak Ising interactions  is equivalent to the \padic~Zabrodin model -- a simple toy model of AdS/CFT. This equivalence enriches our understanding of both models. In particular, the dual electrostatics model is instrumental in illustrating the holographic information properties inherent to the Zabrodin model, and the Zabrodin model's effective boundary action (Eq.~\eqref{eq:Sboundary}) can be leveraged to compute the 3- and 4-point correlation functions of the HFM.   A challenge is presented, however, by the fact that the Ising-fracton hyperbolic model is dual to not one but a large set of copies of the Zabrodin model, and the boundary duals of these theories are all scrambled together.

	Our findings invite future quantitative exploration
	of emerging new concepts. Going beyond the low-energy sector, the series of Zabrodin models that make up the dual of the HFM become interlinked. Each  tree is described using \padic~geometry, yet when they're combined, conventional hyperbolic geometry is restored. 
	This raises intriguing questions around whether this transition of geometry carries a deeper meaning.

	Furthermore, when the fractal trees reduce to geodesics (the equivalent of ``straight lines" in hyperbolic space) in the limit of   $2$-adic geometry \cite{Yan2019PhysRevBfracton1,Yan2019PhysRevBfracton2,Yan2020PhysRevB}, the associated electrostatics theory   represents a classical analogue of a ``bit thread'' \cite{Freedman2017CMaPh,Headrick2018CQG,Chen2018arXiv}. This notion of a ``bit thread'' is a string in the bulk with its endpoints entangled on the boundary. Our work proposes its extension as a ``bit tree'' in the bulk with multiple entangled endpoints. This opens up fascinating possibilities; for instance, could these ``bit trees'' offer fresh insights into the structure of holographic entanglement entropy and holographic code based on tensor networks? We hope that the equivalence demonstrated in our work can serve as a key node, concretely linking various facets of AdS/CFT.
	
	\section*{Acknowledgment}
	H.Y. thanks Hao-Yu Sun and Andriy H. Nevidomskyy for helpful discussions. 
	H.Y. was supported by the U.S. National Science Foundation Division of Materials Research under the Award DMR-1917511. Y.O. is supported by the ISF center of excellence.
	\bibliography{padicReferences} 

\begin{thebibliography}{44}%
\makeatletter
\providecommand \@ifxundefined [1]{%
 \@ifx{#1\undefined}
}%
\providecommand \@ifnum [1]{%
 \ifnum #1\expandafter \@firstoftwo
 \else \expandafter \@secondoftwo
 \fi
}%
\providecommand \@ifx [1]{%
 \ifx #1\expandafter \@firstoftwo
 \else \expandafter \@secondoftwo
 \fi
}%
\providecommand \natexlab [1]{#1}%
\providecommand \enquote  [1]{``#1''}%
\providecommand \bibnamefont  [1]{#1}%
\providecommand \bibfnamefont [1]{#1}%
\providecommand \citenamefont [1]{#1}%
\providecommand \href@noop [0]{\@secondoftwo}%
\providecommand \href [0]{\begingroup \@sanitize@url \@href}%
\providecommand \@href[1]{\@@startlink{#1}\@@href}%
\providecommand \@@href[1]{\endgroup#1\@@endlink}%
\providecommand \@sanitize@url [0]{\catcode `\\12\catcode `\$12\catcode
  `\&12\catcode `\#12\catcode `\^12\catcode `\_12\catcode `\%12\relax}%
\providecommand \@@startlink[1]{}%
\providecommand \@@endlink[0]{}%
\providecommand \url  [0]{\begingroup\@sanitize@url \@url }%
\providecommand \@url [1]{\endgroup\@href {#1}{\urlprefix }}%
\providecommand \urlprefix  [0]{URL }%
\providecommand \Eprint [0]{\href }%
\providecommand \doibase [0]{https://doi.org/}%
\providecommand \selectlanguage [0]{\@gobble}%
\providecommand \bibinfo  [0]{\@secondoftwo}%
\providecommand \bibfield  [0]{\@secondoftwo}%
\providecommand \translation [1]{[#1]}%
\providecommand \BibitemOpen [0]{}%
\providecommand \bibitemStop [0]{}%
\providecommand \bibitemNoStop [0]{.\EOS\space}%
\providecommand \EOS [0]{\spacefactor3000\relax}%
\providecommand \BibitemShut  [1]{\csname bibitem#1\endcsname}%
\let\auto@bib@innerbib\@empty
\bibitem [{\citenamefont {Maldacena}(1999)}]{Maldacena1999}%
  \BibitemOpen
  \bibfield  {author} {\bibinfo {author} {\bibfnamefont {J.}~\bibnamefont
  {Maldacena}},\ }\href {https://doi.org/10.1023/A:1026654312961} {\bibfield
  {journal} {\bibinfo  {journal} {Int. J. Theor. Phys.}\ }\textbf {\bibinfo
  {volume} {38}},\ \bibinfo {pages} {1113} (\bibinfo {year}
  {1999})}\BibitemShut {NoStop}%
\bibitem [{\citenamefont {Witten}(1998)}]{Witten1998}%
  \BibitemOpen
  \bibfield  {author} {\bibinfo {author} {\bibfnamefont {E.}~\bibnamefont
  {Witten}},\ }\href {https://doi.org/10.4310/ATMP.1998.v2.n2.a2} {\bibfield
  {journal} {\bibinfo  {journal} {Adv. Theor. Math. Phys.}\ }\textbf {\bibinfo
  {volume} {2}},\ \bibinfo {pages} {253} (\bibinfo {year} {1998})}\BibitemShut
  {NoStop}%
\bibitem [{\citenamefont {Zaanen}\ \emph {et~al.}(2015)\citenamefont {Zaanen},
  \citenamefont {Liu}, \citenamefont {Sun},\ and\ \citenamefont
  {Schalm}}]{Zaanen2015_ADSCMTbook}%
  \BibitemOpen
  \bibfield  {author} {\bibinfo {author} {\bibfnamefont {J.}~\bibnamefont
  {Zaanen}}, \bibinfo {author} {\bibfnamefont {Y.}~\bibnamefont {Liu}},
  \bibinfo {author} {\bibfnamefont {Y.-W.}\ \bibnamefont {Sun}},\ and\ \bibinfo
  {author} {\bibfnamefont {K.}~\bibnamefont {Schalm}},\ }\href
  {https://doi.org/10.1017/cbo9781139942492} {\emph {\bibinfo {title}
  {Holographic Duality in Condensed Matter Physics}}}\ (\bibinfo  {publisher}
  {Cambridge University Press},\ \bibinfo {year} {2015})\BibitemShut {NoStop}%
\bibitem [{\citenamefont {Hartnoll}\ \emph {et~al.}(2018)\citenamefont
  {Hartnoll}, \citenamefont {Lucas},\ and\ \citenamefont
  {Sachdev}}]{Hartnoll:2018xxg}%
  \BibitemOpen
  \bibfield  {author} {\bibinfo {author} {\bibfnamefont {S.~A.}\ \bibnamefont
  {Hartnoll}}, \bibinfo {author} {\bibfnamefont {A.}~\bibnamefont {Lucas}},\
  and\ \bibinfo {author} {\bibfnamefont {S.}~\bibnamefont {Sachdev}},\ }\href
  {https://mitpress.mit.edu/9780262038430/holographic-quantum-matter/} {\emph
  {\bibinfo {title} {{Holographic Quantum Matter}}}}\ (\bibinfo  {publisher}
  {MIT Press},\ \bibinfo {year} {2018})\BibitemShut {NoStop}%
\bibitem [{\citenamefont {Swingle}(2012)}]{Swingle2012}%
  \BibitemOpen
  \bibfield  {author} {\bibinfo {author} {\bibfnamefont {B.}~\bibnamefont
  {Swingle}},\ }\href {https://doi.org/10.1103/PhysRevD.86.065007} {\bibfield
  {journal} {\bibinfo  {journal} {Phys. Rev. D}\ }\textbf {\bibinfo {volume}
  {86}},\ \bibinfo {pages} {065007} (\bibinfo {year} {2012})}\BibitemShut
  {NoStop}%
\bibitem [{\citenamefont {Pastawski}\ \emph {et~al.}(2015)\citenamefont
  {Pastawski}, \citenamefont {Yoshida}, \citenamefont {Harlow},\ and\
  \citenamefont {Preskill}}]{Pastawski2015}%
  \BibitemOpen
  \bibfield  {author} {\bibinfo {author} {\bibfnamefont {F.}~\bibnamefont
  {Pastawski}}, \bibinfo {author} {\bibfnamefont {B.}~\bibnamefont {Yoshida}},
  \bibinfo {author} {\bibfnamefont {D.}~\bibnamefont {Harlow}},\ and\ \bibinfo
  {author} {\bibfnamefont {J.}~\bibnamefont {Preskill}},\ }\href
  {https://doi.org/10.1007/JHEP06(2015)149} {\bibfield  {journal} {\bibinfo
  {journal} {JHEP}\ }\textbf {\bibinfo {volume} {2015}}\bibinfo  {number} {
  (149)}}\BibitemShut {NoStop}%
\bibitem [{\citenamefont {Yang}\ \emph {et~al.}(2016)\citenamefont {Yang},
  \citenamefont {Hayden},\ and\ \citenamefont {Qi}}]{Yang2016}%
  \BibitemOpen
\bibfield  {number} {  }\bibfield  {author} {\bibinfo {author} {\bibfnamefont
  {Z.}~\bibnamefont {Yang}}, \bibinfo {author} {\bibfnamefont {P.}~\bibnamefont
  {Hayden}},\ and\ \bibinfo {author} {\bibfnamefont {X.-L.}\ \bibnamefont
  {Qi}},\ }\href {https://doi.org/10.1007/jhep01(2016)175} {\bibfield
  {journal} {\bibinfo  {journal} {JHEP}\ }\textbf {\bibinfo {volume}
  {2016}}\bibinfo  {number} { (175)}}\BibitemShut {NoStop}%
\bibitem [{\citenamefont {Chamon}(2005)}]{ChamonPhysRevLett.94.040402}%
  \BibitemOpen
\bibfield  {number} {  }\bibfield  {author} {\bibinfo {author} {\bibfnamefont
  {C.}~\bibnamefont {Chamon}},\ }\href
  {https://doi.org/10.1103/PhysRevLett.94.040402} {\bibfield  {journal}
  {\bibinfo  {journal} {Phys. Rev. Lett.}\ }\textbf {\bibinfo {volume} {94}},\
  \bibinfo {pages} {040402} (\bibinfo {year} {2005})}\BibitemShut {NoStop}%
\bibitem [{\citenamefont {Yoshida}(2013)}]{YoshidaPhysRevB.88.125122}%
  \BibitemOpen
  \bibfield  {author} {\bibinfo {author} {\bibfnamefont {B.}~\bibnamefont
  {Yoshida}},\ }\href {https://doi.org/10.1103/PhysRevB.88.125122} {\bibfield
  {journal} {\bibinfo  {journal} {Phys. Rev. B}\ }\textbf {\bibinfo {volume}
  {88}},\ \bibinfo {pages} {125122} (\bibinfo {year} {2013})}\BibitemShut
  {NoStop}%
\bibitem [{\citenamefont {Bravyi}\ \emph {et~al.}(2011)\citenamefont {Bravyi},
  \citenamefont {Leemhuis},\ and\ \citenamefont {Terhal}}]{BRAVYI2011839}%
  \BibitemOpen
  \bibfield  {author} {\bibinfo {author} {\bibfnamefont {S.}~\bibnamefont
  {Bravyi}}, \bibinfo {author} {\bibfnamefont {B.}~\bibnamefont {Leemhuis}},\
  and\ \bibinfo {author} {\bibfnamefont {B.~M.}\ \bibnamefont {Terhal}},\
  }\href {https://doi.org/https://doi.org/10.1016/j.aop.2010.11.002} {\bibfield
   {journal} {\bibinfo  {journal} {Annals of Physics}\ }\textbf {\bibinfo
  {volume} {326}},\ \bibinfo {pages} {839 } (\bibinfo {year}
  {2011})}\BibitemShut {NoStop}%
\bibitem [{\citenamefont {Haah}(2011)}]{Haah2011}%
  \BibitemOpen
  \bibfield  {author} {\bibinfo {author} {\bibfnamefont {J.}~\bibnamefont
  {Haah}},\ }\href {https://doi.org/10.1103/PhysRevA.83.042330} {\bibfield
  {journal} {\bibinfo  {journal} {Phys. Rev. A}\ }\textbf {\bibinfo {volume}
  {83}},\ \bibinfo {pages} {042330} (\bibinfo {year} {2011})}\BibitemShut
  {NoStop}%
\bibitem [{\citenamefont {Vijay}\ \emph {et~al.}(2015)\citenamefont {Vijay},
  \citenamefont {Haah},\ and\ \citenamefont {Fu}}]{Vijay2015}%
  \BibitemOpen
  \bibfield  {author} {\bibinfo {author} {\bibfnamefont {S.}~\bibnamefont
  {Vijay}}, \bibinfo {author} {\bibfnamefont {J.}~\bibnamefont {Haah}},\ and\
  \bibinfo {author} {\bibfnamefont {L.}~\bibnamefont {Fu}},\ }\href
  {https://doi.org/10.1103/PhysRevB.92.235136} {\bibfield  {journal} {\bibinfo
  {journal} {Phys. Rev. B}\ }\textbf {\bibinfo {volume} {92}},\ \bibinfo
  {pages} {235136} (\bibinfo {year} {2015})}\BibitemShut {NoStop}%
\bibitem [{\citenamefont {Pretko}(2017{\natexlab{a}})}]{Pretko2017a}%
  \BibitemOpen
  \bibfield  {author} {\bibinfo {author} {\bibfnamefont {M.}~\bibnamefont
  {Pretko}},\ }\href {https://doi.org/10.1103/PhysRevB.96.035119} {\bibfield
  {journal} {\bibinfo  {journal} {Phys. Rev. B}\ }\textbf {\bibinfo {volume}
  {96}},\ \bibinfo {pages} {035119} (\bibinfo {year}
  {2017}{\natexlab{a}})}\BibitemShut {NoStop}%
\bibitem [{\citenamefont {Nandkishore}\ and\ \citenamefont
  {Hermele}(2019)}]{Nandkishoreannurev}%
  \BibitemOpen
  \bibfield  {author} {\bibinfo {author} {\bibfnamefont {R.~M.}\ \bibnamefont
  {Nandkishore}}\ and\ \bibinfo {author} {\bibfnamefont {M.}~\bibnamefont
  {Hermele}},\ }\href
  {https://doi.org/10.1146/annurev-conmatphys-031218-013604} {\bibfield
  {journal} {\bibinfo  {journal} {Annual Review of Condensed Matter Physics}\
  }\textbf {\bibinfo {volume} {10}},\ \bibinfo {pages} {295} (\bibinfo {year}
  {2019})}\BibitemShut {NoStop}%
\bibitem [{\citenamefont {Pretko}\ \emph {et~al.}(2020)\citenamefont {Pretko},
  \citenamefont {Chen},\ and\ \citenamefont {You}}]{pretko2020fractonReview}%
  \BibitemOpen
  \bibfield  {author} {\bibinfo {author} {\bibfnamefont {M.}~\bibnamefont
  {Pretko}}, \bibinfo {author} {\bibfnamefont {X.}~\bibnamefont {Chen}},\ and\
  \bibinfo {author} {\bibfnamefont {Y.}~\bibnamefont {You}},\ }\href
  {https://www.worldscientific.com/doi/10.1142/S0217751X20300033} {\bibfield
  {journal} {\bibinfo  {journal} {International Journal of Modern Physics A}\
  }\textbf {\bibinfo {volume} {35}},\ \bibinfo {pages} {2030003} (\bibinfo
  {year} {2020})}\BibitemShut {NoStop}%
\bibitem [{\citenamefont {Gromov}\ and\ \citenamefont
  {Radzihovsky}(2022)}]{gromov2022fractonReview}%
  \BibitemOpen
  \bibfield  {author} {\bibinfo {author} {\bibfnamefont {A.}~\bibnamefont
  {Gromov}}\ and\ \bibinfo {author} {\bibfnamefont {L.}~\bibnamefont
  {Radzihovsky}},\ }\href@noop {} {\bibinfo {title} {Fracton matter}} (\bibinfo
  {year} {2022}),\ \Eprint {https://arxiv.org/abs/2211.05130} {arXiv:2211.05130
  [cond-mat.str-el]} \BibitemShut {NoStop}%
\bibitem [{\citenamefont {Xu}\ and\ \citenamefont {Ho\ifmmode~\check{r}\else
  \v{r}\fi{}ava}(2010)}]{Xu2010PhysRevD}%
  \BibitemOpen
  \bibfield  {author} {\bibinfo {author} {\bibfnamefont {C.}~\bibnamefont
  {Xu}}\ and\ \bibinfo {author} {\bibfnamefont {P.}~\bibnamefont
  {Ho\ifmmode~\check{r}\else \v{r}\fi{}ava}},\ }\href
  {https://doi.org/10.1103/PhysRevD.81.104033} {\bibfield  {journal} {\bibinfo
  {journal} {Phys. Rev. D}\ }\textbf {\bibinfo {volume} {81}},\ \bibinfo
  {pages} {104033} (\bibinfo {year} {2010})}\BibitemShut {NoStop}%
\bibitem [{\citenamefont {Pretko}(2017{\natexlab{b}})}]{Pretko2017PhysRevD}%
  \BibitemOpen
  \bibfield  {author} {\bibinfo {author} {\bibfnamefont {M.}~\bibnamefont
  {Pretko}},\ }\href {https://doi.org/10.1103/PhysRevD.96.024051} {\bibfield
  {journal} {\bibinfo  {journal} {Phys. Rev. D}\ }\textbf {\bibinfo {volume}
  {96}},\ \bibinfo {pages} {024051} (\bibinfo {year}
  {2017}{\natexlab{b}})}\BibitemShut {NoStop}%
\bibitem [{\citenamefont {Yan}(2019{\natexlab{a}})}]{Yan2019PhysRevBfracton1}%
  \BibitemOpen
  \bibfield  {author} {\bibinfo {author} {\bibfnamefont {H.}~\bibnamefont
  {Yan}},\ }\href {https://doi.org/10.1103/PhysRevB.99.155126} {\bibfield
  {journal} {\bibinfo  {journal} {Phys. Rev. B}\ }\textbf {\bibinfo {volume}
  {99}},\ \bibinfo {pages} {155126} (\bibinfo {year}
  {2019}{\natexlab{a}})}\BibitemShut {NoStop}%
\bibitem [{\citenamefont {Yan}(2019{\natexlab{b}})}]{Yan2019PhysRevBfracton2}%
  \BibitemOpen
  \bibfield  {author} {\bibinfo {author} {\bibfnamefont {H.}~\bibnamefont
  {Yan}},\ }\href {https://doi.org/10.1103/PhysRevB.100.245138} {\bibfield
  {journal} {\bibinfo  {journal} {Phys. Rev. B}\ }\textbf {\bibinfo {volume}
  {100}},\ \bibinfo {pages} {245138} (\bibinfo {year}
  {2019}{\natexlab{b}})}\BibitemShut {NoStop}%
\bibitem [{\citenamefont {Yan}(2020)}]{Yan2020PhysRevB}%
  \BibitemOpen
  \bibfield  {author} {\bibinfo {author} {\bibfnamefont {H.}~\bibnamefont
  {Yan}},\ }\href {https://doi.org/10.1103/PhysRevB.102.161119} {\bibfield
  {journal} {\bibinfo  {journal} {Phys. Rev. B}\ }\textbf {\bibinfo {volume}
  {102}},\ \bibinfo {pages} {161119} (\bibinfo {year} {2020})}\BibitemShut
  {NoStop}%
\bibitem [{\citenamefont {Jahn}\ \emph {et~al.}(2019)\citenamefont {Jahn},
  \citenamefont {Gluza}, \citenamefont {Pastawski},\ and\ \citenamefont
  {Eisert}}]{Jahneaaw0092}%
  \BibitemOpen
  \bibfield  {author} {\bibinfo {author} {\bibfnamefont {A.}~\bibnamefont
  {Jahn}}, \bibinfo {author} {\bibfnamefont {M.}~\bibnamefont {Gluza}},
  \bibinfo {author} {\bibfnamefont {F.}~\bibnamefont {Pastawski}},\ and\
  \bibinfo {author} {\bibfnamefont {J.}~\bibnamefont {Eisert}},\ }\bibfield
  {journal} {\bibinfo  {journal} {Science Advances}\ }\textbf {\bibinfo
  {volume} {5}},\ \href {https://doi.org/10.1126/sciadv.aaw0092}
  {10.1126/sciadv.aaw0092} (\bibinfo {year} {2019})\BibitemShut {NoStop}%
\bibitem [{\citenamefont {Zabrodin}(1989)}]{zabrodin1989non}%
  \BibitemOpen
  \bibfield  {author} {\bibinfo {author} {\bibfnamefont {A.~V.}\ \bibnamefont
  {Zabrodin}},\ }\href {https://doi.org/10.1007/bf01238811} {\bibfield
  {journal} {\bibinfo  {journal} {Commun. Math. Phys.}\ }\textbf {\bibinfo
  {volume} {123}},\ \bibinfo {pages} {463} (\bibinfo {year}
  {1989})}\BibitemShut {NoStop}%
\bibitem [{\citenamefont {Breuckmann}\ \emph {et~al.}(2020)\citenamefont
  {Breuckmann}, \citenamefont {Placke},\ and\ \citenamefont
  {Roy}}]{breuckmann2020critical}%
  \BibitemOpen
  \bibfield  {author} {\bibinfo {author} {\bibfnamefont {N.~P.}\ \bibnamefont
  {Breuckmann}}, \bibinfo {author} {\bibfnamefont {B.}~\bibnamefont {Placke}},\
  and\ \bibinfo {author} {\bibfnamefont {A.}~\bibnamefont {Roy}},\ }\href@noop
  {} {\bibfield  {journal} {\bibinfo  {journal} {Physical Review E}\ }\textbf
  {\bibinfo {volume} {101}},\ \bibinfo {pages} {022124} (\bibinfo {year}
  {2020})}\BibitemShut {NoStop}%
\bibitem [{\citenamefont {Koblitz}(1977)}]{koblitz}%
  \BibitemOpen
  \bibfield  {author} {\bibinfo {author} {\bibfnamefont {N.}~\bibnamefont
  {Koblitz}},\ }\href
  {https://link.springer.com/book/10.1007/978-1-4612-1112-9} {\emph {\bibinfo
  {title} {p-adic Numbers, p-adic Analysis, and Zeta-Functions}}}\ (\bibinfo
  {publisher} {Springer},\ \bibinfo {year} {1977})\BibitemShut {NoStop}%
\bibitem [{\citenamefont {Gouv{\^e}a}(2020)}]{gouvea}%
  \BibitemOpen
  \bibfield  {author} {\bibinfo {author} {\bibfnamefont {F.~Q.}\ \bibnamefont
  {Gouv{\^e}a}},\ }\href {https://doi.org/10.1007/978-3-030-47295-5} {\emph
  {\bibinfo {title} {p-adic Numbers: An Introduction}}}\ (\bibinfo  {publisher}
  {Springer},\ \bibinfo {year} {2020})\BibitemShut {NoStop}%
\bibitem [{\citenamefont {Freund}\ and\ \citenamefont
  {Olson}(1987)}]{freund1987non}%
  \BibitemOpen
  \bibfield  {author} {\bibinfo {author} {\bibfnamefont {P.~G.}\ \bibnamefont
  {Freund}}\ and\ \bibinfo {author} {\bibfnamefont {M.}~\bibnamefont {Olson}},\
  }\href@noop {} {\bibfield  {journal} {\bibinfo  {journal} {Physics Letters
  B}\ }\textbf {\bibinfo {volume} {199}},\ \bibinfo {pages} {186} (\bibinfo
  {year} {1987})}\BibitemShut {NoStop}%
\bibitem [{\citenamefont {Spokoiny}(1988)}]{spokoiny1988quantum}%
  \BibitemOpen
  \bibfield  {author} {\bibinfo {author} {\bibfnamefont {B.~L.}\ \bibnamefont
  {Spokoiny}},\ }\href
  {https://doi.org/https://doi.org/10.1016/0370-2693(88)90637-5} {\bibfield
  {journal} {\bibinfo  {journal} {Phys. Lett. B}\ }\textbf {\bibinfo {volume}
  {208}},\ \bibinfo {pages} {401} (\bibinfo {year} {1988})}\BibitemShut
  {NoStop}%
\bibitem [{\citenamefont {Zhang}(1988)}]{zhang1988lagrangian}%
  \BibitemOpen
  \bibfield  {author} {\bibinfo {author} {\bibfnamefont {R.}~\bibnamefont
  {Zhang}},\ }\href
  {https://doi.org/https://doi.org/10.1016/0370-2693(88)90937-9} {\bibfield
  {journal} {\bibinfo  {journal} {Phys. Lett. B}\ }\textbf {\bibinfo {volume}
  {209}},\ \bibinfo {pages} {229} (\bibinfo {year} {1988})}\BibitemShut
  {NoStop}%
\bibitem [{\citenamefont {Gubser}\ \emph {et~al.}(2017)\citenamefont {Gubser},
  \citenamefont {Knaute}, \citenamefont {Parikh}, \citenamefont {Samberg},\
  and\ \citenamefont {Witaszczyk}}]{gubser2017p}%
  \BibitemOpen
  \bibfield  {author} {\bibinfo {author} {\bibfnamefont {S.~S.}\ \bibnamefont
  {Gubser}}, \bibinfo {author} {\bibfnamefont {J.}~\bibnamefont {Knaute}},
  \bibinfo {author} {\bibfnamefont {S.}~\bibnamefont {Parikh}}, \bibinfo
  {author} {\bibfnamefont {A.}~\bibnamefont {Samberg}},\ and\ \bibinfo {author}
  {\bibfnamefont {P.}~\bibnamefont {Witaszczyk}},\ }\href
  {https://doi.org/10.1007/s00220-016-2813-6} {\bibfield  {journal} {\bibinfo
  {journal} {Commu. Math. Phys.}\ }\textbf {\bibinfo {volume} {352}},\ \bibinfo
  {pages} {1019} (\bibinfo {year} {2017})}\BibitemShut {NoStop}%
\bibitem [{\citenamefont {Heydeman}\ \emph {et~al.}(2017)\citenamefont
  {Heydeman}, \citenamefont {Marcolli}, \citenamefont {Saberi},\ and\
  \citenamefont {Stoica}}]{heydeman2016tensor}%
  \BibitemOpen
  \bibfield  {author} {\bibinfo {author} {\bibfnamefont {M.}~\bibnamefont
  {Heydeman}}, \bibinfo {author} {\bibfnamefont {M.}~\bibnamefont {Marcolli}},
  \bibinfo {author} {\bibfnamefont {I.}~\bibnamefont {Saberi}},\ and\ \bibinfo
  {author} {\bibfnamefont {B.}~\bibnamefont {Stoica}},\ }\href@noop {}
  {\bibinfo {title} {Tensor networks, $p$-adic fields, and algebraic curves:
  arithmetic and the ads$_3$/cft$_2$ correspondence}} (\bibinfo {year}
  {2017}),\ \Eprint {https://arxiv.org/abs/1605.07639} {arXiv:1605.07639
  [hep-th]} \BibitemShut {NoStop}%
\bibitem [{\citenamefont {Yan}\ \emph {et~al.}(2022)\citenamefont {Yan},
  \citenamefont {Slagle},\ and\ \citenamefont {Nevidomskyy}}]{yan2022ycube}%
  \BibitemOpen
  \bibfield  {author} {\bibinfo {author} {\bibfnamefont {H.}~\bibnamefont
  {Yan}}, \bibinfo {author} {\bibfnamefont {K.}~\bibnamefont {Slagle}},\ and\
  \bibinfo {author} {\bibfnamefont {A.~H.}\ \bibnamefont {Nevidomskyy}},\
  }\href@noop {} {\bibinfo {title} {Y-cube model and fractal structure of
  subdimensional particles on hyperbolic lattices}} (\bibinfo {year} {2022}),\
  \Eprint {https://arxiv.org/abs/2211.15829} {arXiv:2211.15829 [quant-ph]}
  \BibitemShut {NoStop}%
\bibitem [{\citenamefont {Manin}\ and\ \citenamefont
  {Marcolli}(2002)}]{manin2002holography}%
  \BibitemOpen
  \bibfield  {author} {\bibinfo {author} {\bibfnamefont {Y.~I.}\ \bibnamefont
  {Manin}}\ and\ \bibinfo {author} {\bibfnamefont {M.}~\bibnamefont
  {Marcolli}},\ }\href@noop {} {\bibinfo {title} {Holography principle and
  arithmetic of algebraic curves}} (\bibinfo {year} {2002}),\ \Eprint
  {https://arxiv.org/abs/hep-th/0201036} {arXiv:hep-th/0201036 [hep-th]}
  \BibitemShut {NoStop}%
\bibitem [{\citenamefont {Ba\~nados}\ \emph {et~al.}(1992)\citenamefont
  {Ba\~nados}, \citenamefont {Teitelboim},\ and\ \citenamefont
  {Zanelli}}]{banados1992black}%
  \BibitemOpen
  \bibfield  {author} {\bibinfo {author} {\bibfnamefont {M.}~\bibnamefont
  {Ba\~nados}}, \bibinfo {author} {\bibfnamefont {C.}~\bibnamefont
  {Teitelboim}},\ and\ \bibinfo {author} {\bibfnamefont {J.}~\bibnamefont
  {Zanelli}},\ }\href {https://doi.org/10.1103/PhysRevLett.69.1849} {\bibfield
  {journal} {\bibinfo  {journal} {Phys. Rev. Lett.}\ }\textbf {\bibinfo
  {volume} {69}},\ \bibinfo {pages} {1849} (\bibinfo {year}
  {1992})}\BibitemShut {NoStop}%
\bibitem [{\citenamefont {Steif}(1996)}]{steif1996supergeometry}%
  \BibitemOpen
  \bibfield  {author} {\bibinfo {author} {\bibfnamefont {A.~R.}\ \bibnamefont
  {Steif}},\ }\href {https://doi.org/10.1103/PhysRevD.53.5521} {\bibfield
  {journal} {\bibinfo  {journal} {Phys. Rev. D}\ }\textbf {\bibinfo {volume}
  {53}},\ \bibinfo {pages} {5521} (\bibinfo {year} {1996})}\BibitemShut
  {NoStop}%
\bibitem [{\citenamefont {Heydeman}\ \emph {et~al.}(2018)\citenamefont
  {Heydeman}, \citenamefont {Marcolli}, \citenamefont {Parikh},\ and\
  \citenamefont {Saberi}}]{heydeman2018nonarchimedean}%
  \BibitemOpen
  \bibfield  {author} {\bibinfo {author} {\bibfnamefont {M.}~\bibnamefont
  {Heydeman}}, \bibinfo {author} {\bibfnamefont {M.}~\bibnamefont {Marcolli}},
  \bibinfo {author} {\bibfnamefont {S.}~\bibnamefont {Parikh}},\ and\ \bibinfo
  {author} {\bibfnamefont {I.}~\bibnamefont {Saberi}},\ }\href@noop {}
  {\bibinfo {title} {Nonarchimedean holographic entropy from networks of
  perfect tensors}} (\bibinfo {year} {2018}),\ \Eprint
  {https://arxiv.org/abs/1812.04057} {arXiv:1812.04057 [hep-th]} \BibitemShut
  {NoStop}%
\bibitem [{\citenamefont {Hung}\ \emph {et~al.}(2019)\citenamefont {Hung},
  \citenamefont {Li},\ and\ \citenamefont {Melby-Thompson}}]{hung2019p}%
  \BibitemOpen
  \bibfield  {author} {\bibinfo {author} {\bibfnamefont {L.-Y.}\ \bibnamefont
  {Hung}}, \bibinfo {author} {\bibfnamefont {W.}~\bibnamefont {Li}},\ and\
  \bibinfo {author} {\bibfnamefont {C.~M.}\ \bibnamefont {Melby-Thompson}},\
  }\href {https://doi.org/10.1007/jhep04(2019)170} {\bibfield  {journal}
  {\bibinfo  {journal} {JHEP}\ }\textbf {\bibinfo {volume} {2019}}\bibinfo
  {number} { (170)}}\BibitemShut {NoStop}%
\bibitem [{\citenamefont {Ebert}\ \emph {et~al.}(2022)\citenamefont {Ebert},
  \citenamefont {Sun},\ and\ \citenamefont {Zhang}}]{ebert2019probing}%
  \BibitemOpen
\bibfield  {number} {  }\bibfield  {author} {\bibinfo {author} {\bibfnamefont
  {S.}~\bibnamefont {Ebert}}, \bibinfo {author} {\bibfnamefont {H.-Y.}\
  \bibnamefont {Sun}},\ and\ \bibinfo {author} {\bibfnamefont {M.-Y.}\
  \bibnamefont {Zhang}},\ }\href@noop {} {\bibinfo {title} {Probing holography
  in $p$-adic cft}} (\bibinfo {year} {2022}),\ \Eprint
  {https://arxiv.org/abs/1911.06313} {arXiv:1911.06313 [hep-th]} \BibitemShut
  {NoStop}%
\bibitem [{\citenamefont {Chen}\ \emph
  {et~al.}(2021{\natexlab{a}})\citenamefont {Chen}, \citenamefont {Liu},\ and\
  \citenamefont {Hung}}]{chen2021bending1}%
  \BibitemOpen
  \bibfield  {author} {\bibinfo {author} {\bibfnamefont {L.}~\bibnamefont
  {Chen}}, \bibinfo {author} {\bibfnamefont {X.}~\bibnamefont {Liu}},\ and\
  \bibinfo {author} {\bibfnamefont {L.-Y.}\ \bibnamefont {Hung}},\ }\href
  {https://doi.org/10.1007/JHEP06(2021)094} {\bibfield  {journal} {\bibinfo
  {journal} {JHEP}\ }\textbf {\bibinfo {volume} {2021}}\bibinfo  {number} {
  (94)}}\BibitemShut {NoStop}%
\bibitem [{\citenamefont {Chen}\ \emph
  {et~al.}(2021{\natexlab{b}})\citenamefont {Chen}, \citenamefont {Liu},\ and\
  \citenamefont {Hung}}]{chen2021bending2}%
  \BibitemOpen
\bibfield  {number} {  }\bibfield  {author} {\bibinfo {author} {\bibfnamefont
  {L.}~\bibnamefont {Chen}}, \bibinfo {author} {\bibfnamefont {X.}~\bibnamefont
  {Liu}},\ and\ \bibinfo {author} {\bibfnamefont {L.-Y.}\ \bibnamefont
  {Hung}},\ }\href {https://doi.org/10.1007/JHEP09(2021)097} {\bibfield
  {journal} {\bibinfo  {journal} {JHEP}\ }\textbf {\bibinfo {volume}
  {2021}}\bibinfo  {number} { (97)}}\BibitemShut {NoStop}%
\bibitem [{\citenamefont {{Freedman}}\ and\ \citenamefont
  {{Headrick}}(2017)}]{Freedman2017CMaPh}%
  \BibitemOpen
\bibfield  {number} {  }\bibfield  {author} {\bibinfo {author} {\bibfnamefont
  {M.}~\bibnamefont {{Freedman}}}\ and\ \bibinfo {author} {\bibfnamefont
  {M.}~\bibnamefont {{Headrick}}},\ }\href
  {https://doi.org/10.1007/s00220-016-2796-3} {\bibfield  {journal} {\bibinfo
  {journal} {Commun. Math. Phys}\ }\textbf {\bibinfo {volume} {352}},\ \bibinfo
  {pages} {407} (\bibinfo {year} {2017})}\BibitemShut {NoStop}%
\bibitem [{\citenamefont {{Headrick}}\ and\ \citenamefont
  {{Hubeny}}(2018)}]{Headrick2018CQG}%
  \BibitemOpen
  \bibfield  {author} {\bibinfo {author} {\bibfnamefont {M.}~\bibnamefont
  {{Headrick}}}\ and\ \bibinfo {author} {\bibfnamefont {V.~E.}\ \bibnamefont
  {{Hubeny}}},\ }\href {https://doi.org/10.1088/1361-6382/aab83c} {\bibfield
  {journal} {\bibinfo  {journal} {Classical and Quantum Gravity}\ }\textbf
  {\bibinfo {volume} {35}},\ \bibinfo {eid} {105012} (\bibinfo {year}
  {2018})}\BibitemShut {NoStop}%
\bibitem [{\citenamefont {{Chen}}\ \emph {et~al.}(2018)\citenamefont {{Chen}},
  \citenamefont {{Shu}},\ and\ \citenamefont {{Wu}}}]{Chen2018arXiv}%
  \BibitemOpen
  \bibfield  {author} {\bibinfo {author} {\bibfnamefont {C.-B.}\ \bibnamefont
  {{Chen}}}, \bibinfo {author} {\bibfnamefont {F.-W.}\ \bibnamefont {{Shu}}},\
  and\ \bibinfo {author} {\bibfnamefont {M.-H.}\ \bibnamefont {{Wu}}},\
  }\href@noop {} {\bibfield  {journal} {\bibinfo  {journal} {ArXiv e-prints}\ }
  (\bibinfo {year} {2018})},\ \Eprint {https://arxiv.org/abs/1804.00441}
  {arXiv:1804.00441 [hep-th]} \BibitemShut {NoStop}%
\bibitem [{\citenamefont {Basteiro}\ \emph {et~al.}(2022)\citenamefont
  {Basteiro}, \citenamefont {Giulio}, \citenamefont {Erdmenger}, \citenamefont
  {Karl}, \citenamefont {Meyer},\ and\ \citenamefont
  {Xian}}]{basteiro2022towards}%
  \BibitemOpen
  \bibfield  {author} {\bibinfo {author} {\bibfnamefont {P.}~\bibnamefont
  {Basteiro}}, \bibinfo {author} {\bibfnamefont {G.~D.}\ \bibnamefont
  {Giulio}}, \bibinfo {author} {\bibfnamefont {J.}~\bibnamefont {Erdmenger}},
  \bibinfo {author} {\bibfnamefont {J.}~\bibnamefont {Karl}}, \bibinfo {author}
  {\bibfnamefont {R.}~\bibnamefont {Meyer}},\ and\ \bibinfo {author}
  {\bibfnamefont {Z.-Y.}\ \bibnamefont {Xian}},\ }\href
  {https://doi.org/10.21468/SciPostPhys.13.5.103} {\bibfield  {journal}
  {\bibinfo  {journal} {SciPost Phys.}\ }\textbf {\bibinfo {volume} {13}},\
  \bibinfo {pages} {103} (\bibinfo {year} {2022})}\BibitemShut {NoStop}%
\end{thebibliography}%
	
	\clearpage 
	\setcounter{equation}{0}
	\setcounter{figure}{0}
	\setcounter{table}{0}
	\makeatletter
	\renewcommand{\theequation}{S\arabic{equation}}
	\renewcommand{\thefigure}{S\arabic{figure}}
	\renewcommand{\bibnumfmt}[1]{[#1]}
	\renewcommand{\citenumfont}[1]{#1}
	
	\onecolumngrid
	 
	\begin{center}
		\large{\textbf{Appendix for ``$p$-adic Holography from Hyperbolic Fracton Model"}}
	\end{center}

	\section{Counting the DOF}
	\label{SEC_APP_count_DOF}
	
	Let us discuss the counting of DOFs when we switch from the original HFM to the dual one living on the edges/vertices.
	The edge terms in $\mathcal{H}_{\text{Ising}}$ can be thought of as living on all the fractal trees of coordination number $n/2=(p+1)$ that span the hyperbolic lattice, with two trees intersecting at each vertex of the lattice. The vertex terms in $\mathcal{H}_{\text{HFM}}$ couple the trees together, but in the limit where $U$ tends to infinity in units of inverse temperature, the trees decouple, and the model $\mathcal{H}_{\text{HFM-Ising}}$ becomes equivalent to a decoupled set of Zabrodin-models. To see how the degrees of freedom align, consider a finite, cut-off version of the hyperbolic lattice with $N_e$ edges, $N_v$ vertices, and $N_p$ plaquettes. For each plaquette, we have one variable $Z_{p_i}$, but in the $U\rightarrow \infty$ limit, we also have one rigid constraint for each vertex. Additionally, $\mathcal{H}_{\text{HFM-Ising}}$ is invariant if we send every plaquette variable to its negative: $Z_{p_i} \rightarrow -Z_{p_i}$. In total we are left with $N_p-N_v-1$ degrees of freedom. In the tree-description, we have one variable ${\bf E}_{e_i}$ for each edge. There are more edges than vertices on the lattice, but in the tree-description, each vertex term imposes a rigid constraint on \textit{two} intersecting trees, leaving us with $N_e-2N_v$ degrees of freedom. Including also the external face in counting the total number of faces $N_f = N_p + 1$, the equality between the degrees of freedom in the two descriptions amounts to Euler's formula for connected plane graphs: $N_v-N_e+N_f=2$.

	\section{Rindler reconstruction}
	\label{SEC_APP_Rindler}
	
	The fractal-tree electrostatics model, dual to an HFM subsector, helps the demonstration of holographic properties. A notable example is the Rindler reconstruction, asserting that part of the boundary data can reconstruct the bulk within the wedge of the minimal surface, as mentioned in Fig.~\ref{fig_rindler_1panel}.
	
\begin{figure}[ht!]
		\centering
		\includegraphics[width=0.8\columnwidth]{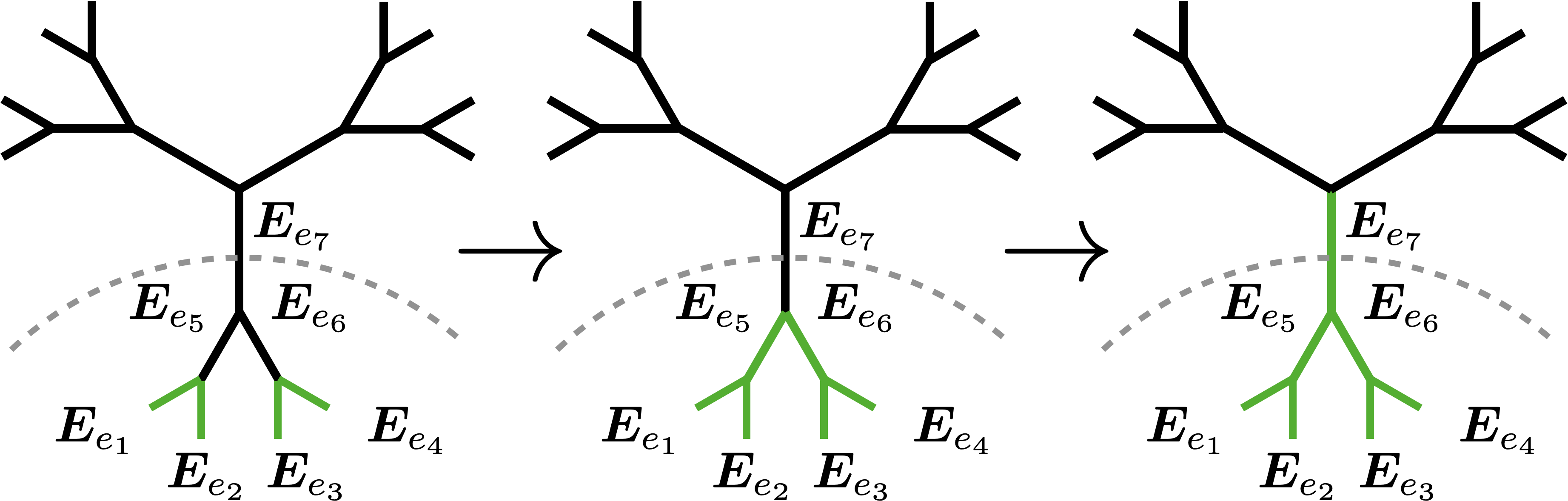}
		\caption{\label{fig_rindler}  
			AdS-Rindler reconstruction of the electrostatics theory on the fractal tree.
		} 
	\end{figure}

	Let's explain in more detail the reconstruction using  Fig.~\ref{fig_rindler}. Here, the known boundary electric fields are denoted as $\bm{E}_{e_{1,2,3,4}}$. The aim of the Rindler reconstruction is to determine which bulk electric fields can be reconstructed.

	In the low-energy limit, the charge-free condition imposed on each vertex leads to:
	\[ 
	\sum_{i=1,2,5}
	\bm{E}_{e_{i}} \cdot \hat{\bm{r}}_{e_{i}}   = 0, \quad 
	\sum_{i=3,4,6}
	\bm{E}_{e_{i}} \cdot \hat{\bm{r}}_{e_{i}}   = 0, 
	\]
	Using these equations, we can reconstruct the values of $\bm{E}_{e_5}$ and $\bm{E}_{e_6}$ from $\bm{E}_{e_{1,2,3,4}}$, which further allows us to determine $\bm{E}_{e_7}$.
	However, the reconstruction process stops beyond $\bm{E}_{e_{7}}$. Beyond this point, we encounter two edges with unknown electric fields, halting further reconstruction. It's important to note that $\bm{E}_{e_{7}} $ represents the minimal ``surface" in the bulk that separates the known boundary segment from the unknown. This demonstrates the Rindler reconstruction on a fractal tree.\\

	\section{Scaling of the lattice}
	\label{SEC_APP_scaling} 
	Let us explain the derivation of Eq.~\ref{EQN_BH_DOF}, a key result in our work.

	\begin{figure}[ht!]
		\centering    \includegraphics[width=0.7\columnwidth]{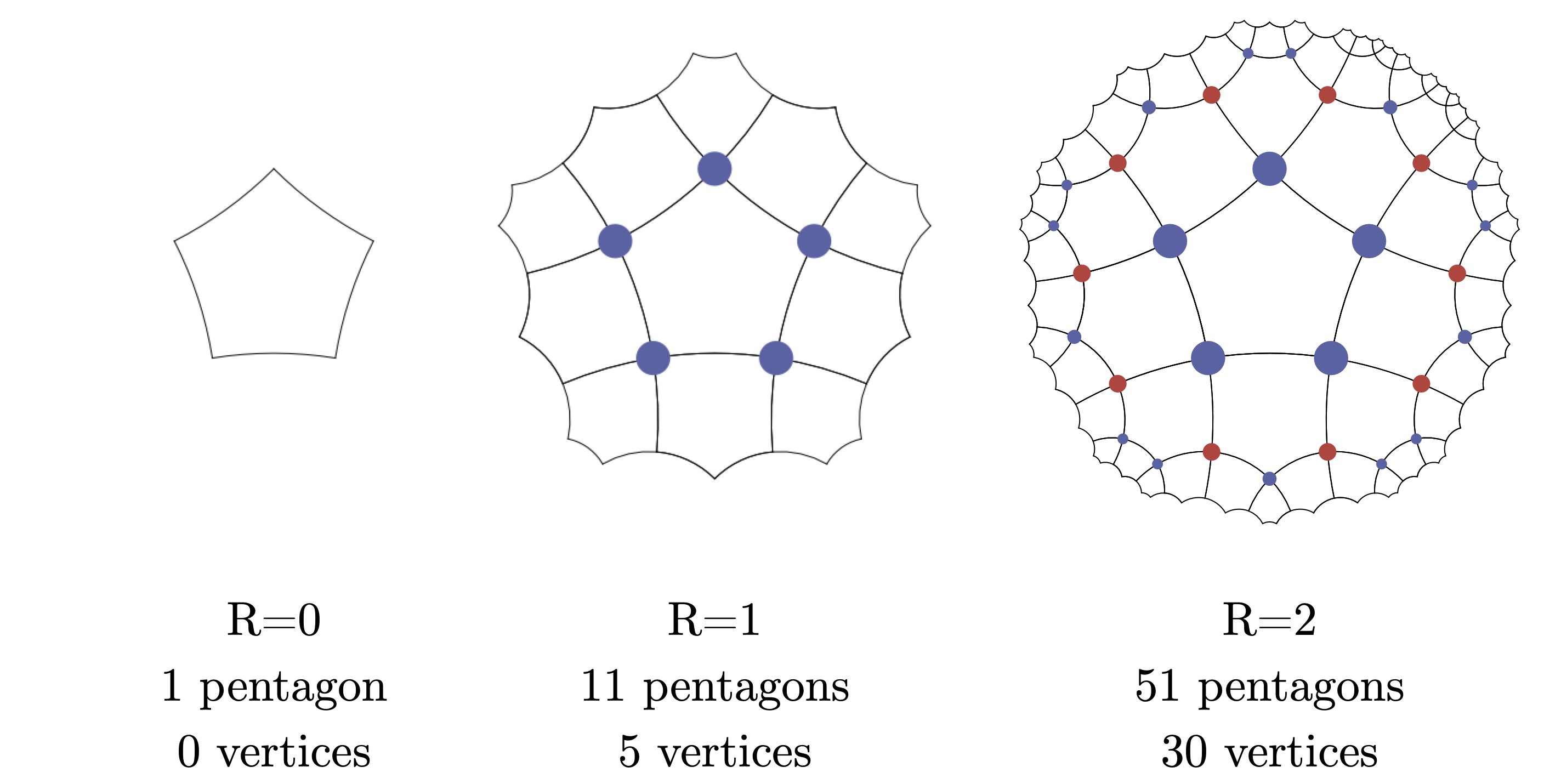}
		\caption{\label{inflationRule}
			Inflation rule for expanding the hyperbolic lattice with Schl\"afli symbol (5,4). 
		} 
	\end{figure}

	To investigate how the number of degrees of freedom and subsystem symmetries scale with system size, we can count the number of vertices and polygons in finite versions of the tessellations with Schl\"afli symbols $(m,n)$. This is a straightforward exercise, using the technology of Ref.~\cite{basteiro2022towards}. The basic idea is to imagine that we generate a tessellation by iteratively performing inflation steps. At each step we add another of layer of polygons as in figure \ref{inflationRule}. In order to perform the counting of vertices and polygons, it is useful to distinguish between two types of vertices, $a$ and $b$, coloured respectively in blue and red in figure \ref{inflationRule}. A red vertex sits at the boundary between two polygons of the previous layer, while a blue vertex touches only one polygon in the previous generation. If we introduce two-component vectors given by 
	\begin{align}
		\vec{v}(R) = 
		\left(
		\begin{matrix}
			\# \text{ of blue vertices in layer $R$}
			\\
			\# \text{ of red vertices in layer $R$}
		\end{matrix}
		\right)\,,
	\end{align}
	then these vectors are given by
	\begin{align}
		\vec{v}(1) = 
		\left(
		\begin{matrix}
			m
			\\
			0
		\end{matrix}
		\right)\,,
		\hspace{20mm}
		\vec{v}(R+1)=M \vec{v}(R)
		\,,
	\end{align}
	where the matrix relating the number of vertex types in neighbouring layers, assuming $m>3$, is given by
	\begin{align}
		M = \left(
		\begin{matrix}
			m-4 + (m-3)(n-3) &&& m-4 + (m-3)(n-4)
			\\
			n-2 &&& n-3
		\end{matrix}
		\right)\,.
	\end{align}
	This matrix has the eigenvalues
	\begin{align}
		\lambda_{\pm}=\frac{PQ-2}{2}
		\pm \frac{1}{2}\sqrt{PQ(PQ-4)}\,,
	\end{align}
	where we have introduced the definitions $P\equiv m-2$ and $Q\equiv n-2$ to make the expressions a bit more compact. By decomposing $\vec{v}(1)$ into eigenvectors of $M$, it becomes easy to compute $\vec{v}(R)$ for arbitrary $R$. By summing over the number of vertices in each layer, we find that the total number of internal vertices $N_v(R)$ in a finite tessellation generated by $R$ inflation steps is given by
	\begin{align}
		N_v(R)
		=\,&
		\sum_{i=1}^R\Big(v_1(i)+v_2(i)\Big)
		\\
		=\,&\frac{P+2}{2}
		\Bigg(
		\frac{\lambda_-^R-1}{\lambda_--1}
		\bigg(1-\sqrt{\frac{PQ}{PQ-4}}\bigg)
		+
		\frac{\lambda_+^R-1}{\lambda_+-1}
		\bigg(1+\sqrt{\frac{PQ}{PQ-4}}\bigg)
		\Bigg)\,.
	\end{align}
	Using the fact that each blue vertex gives rise to $n-2$ polygons in the next layer, and each red vertex gives rise to $n-3$ polygons in the next layer, we can add together the polygons in each layer to determine the number of polygons $N_p(R)$ in a finite tessellation generated by $R$ inflation steps to equal
	\begin{align}
		N_p(R)
		=\,&
		1+\sum_{i=1}^R\bigg((n-2)v_1(i)+(n-3)v_2(i)\bigg)
		\\
		=\,&
		1+\frac{P+2}{2}
		\Bigg(
		\frac{\lambda_-^R-1}{\lambda_--1}
		\bigg(Q-(PQ-2)\sqrt{\frac{Q}{P(PQ-4)}}\bigg)
		+
		\frac{\lambda_+^R-1}{\lambda_+-1}
		\bigg(Q+(PQ-2)\sqrt{\frac{Q}{P(PQ-4)}}\bigg)
		\Bigg)\,.
	\end{align}
	In the limit as the number of inflation steps tends to infinity, the ratio of polygons to vertices tends to a finite number:
	\begin{align}
		\lim_{R\rightarrow \infty}
		\frac{N_p(R)}{N_v(R)}
		=\frac{1}{2}\left(
		Q+\sqrt{\frac{Q(PQ-4)}{P}}
		\right)\,.
	\end{align}
	We can also use the eigendecomposition of $M$ to count the number of trees in a finite hyperbolic graph. For it can be checked that the total number $N_t(n)$ of trees equals the total number of blue vertices, wherefore $N_t(R)$ is given by
	\begin{align}\label{EQN_tree_finite_count}
		N_t(R) = \sum_{i=1}^R v_1(i) = 
		\frac{P+2}{2}
		\Bigg(
		\frac{\lambda_-^R-1}{\lambda_--1}
		\bigg(
		1-(P-2)\sqrt{\frac{Q}{P(PQ-4)}}
		\bigg)
		+
		\frac{\lambda_+^R-1}{\lambda_+-1}
		\bigg(
		1+(P-2)\sqrt{\frac{Q}{P(PQ-4)}}
		\bigg)
		\Bigg)\,.
	\end{align}
	This condenses to the result of  Eq.~\eqref{EQN_BH_DOF}.

\end{document}